\documentclass[aps,
reprint,
prstab,
superscriptaddress,
]{revtex4-1}

\usepackage{siunitx}
\DeclareSIUnit \pC{\pico\coulomb}
\DeclareSIUnit \mrad{\milli\radian}
\DeclareSIUnit \rad{\radian}
\DeclareSIUnit \um{\micro\metre}
\DeclareSIUnit \cm{\centi\metre}
\DeclareSIUnit \nJ{\nano\joule}
\usepackage{amssymb}
\usepackage{amsmath}
\usepackage{graphicx}
\usepackage{color,soul}
\usepackage{dcolumn}% Align table columns on decimal point
\usepackage{hyperref}% add hypertext capabilities
\usepackage[colorinlistoftodos,prependcaption,textsize=tiny]{todonotes}

\begin{document}
%\setcitestyle{super}
%\preprint{APS/123-QED}

\title{Multi-octave high-dynamic range optical spectrometer for single-pulse diagnostic applications}

%\title{Multi-octave optical-spectrometer for single-shot, high-dynamic range and low-light diagnostic applications}
%\title{A Single-shot, ultra-broadband and highly sensitive spectrometer\\for short pulse, high resolution and low light diagnostic applications}% Force line breaks with \\
%\thanks{A footnote to the article title}%

\author{Omid Zarini}
\affiliation{Helmholtz-Zentrum Dresden--Rossendorf, Bautzner Landstrasse 400, 01328 Dresden, Germany}
\affiliation{Technische Universit\"at Dresden, 01069 Dresden, Germany}%
\author{Jurjen Couperus Cabada\u{g}}
\affiliation{Helmholtz-Zentrum Dresden--Rossendorf, Bautzner Landstrasse 400, 01328 Dresden, Germany}
\author{Yen-Yu Chang}
\affiliation{Helmholtz-Zentrum Dresden--Rossendorf, Bautzner Landstrasse 400, 01328 Dresden, Germany}
\author{Alexander K\"ohler}
\affiliation{Helmholtz-Zentrum Dresden--Rossendorf, Bautzner Landstrasse 400, 01328 Dresden, Germany}
\author{Thomas Kurz}
\affiliation{Helmholtz-Zentrum Dresden--Rossendorf, Bautzner Landstrasse 400, 01328 Dresden, Germany}
\affiliation{Technische Universit\"at Dresden, 01069 Dresden, Germany}
\author{Susanne Sch\"obel}
\affiliation{Helmholtz-Zentrum Dresden--Rossendorf, Bautzner Landstrasse 400, 01328 Dresden, Germany}
\affiliation{Technische Universit\"at Dresden, 01069 Dresden, Germany}
\author{Wolfgang Seidel}
\affiliation{Helmholtz-Zentrum Dresden--Rossendorf, Bautzner Landstrasse 400, 01328 Dresden, Germany}
\author{Michael Bussmann}
\affiliation{Helmholtz-Zentrum Dresden--Rossendorf, Bautzner Landstrasse 400, 01328 Dresden, Germany}
\author{Ulrich Schramm}%
\affiliation{Helmholtz-Zentrum Dresden--Rossendorf, Bautzner Landstrasse 400, 01328 Dresden, Germany}
\affiliation{Technische Universit\"at Dresden, 01069 Dresden, Germany}
\author{Arie Irman}%
\email{A.Irman@hzdr.de}
\affiliation{Helmholtz-Zentrum Dresden--Rossendorf, Bautzner Landstrasse 400, 01328 Dresden, Germany}
\author{Alexander Debus}
\email{A.Debus@hzdr.de}
\affiliation{Helmholtz-Zentrum Dresden--Rossendorf, Bautzner Landstrasse 400, 01328 Dresden, Germany}

\date{Draft created on \today}% It is always \today, today,
             %  but any date may be explicitly specified

\begin{abstract}
We present design and realization of an ultra-broadband optical spectrometer capable of measuring the spectral intensity of multi-octave-spanning light sources on a single-pulse basis with a dynamic range of up to 8 orders of magnitude. The instrument is optimized for the characterization of the temporal structure of femtosecond long electron bunches by analyzing the emitted coherent transition radiation (CTR) spectra. The spectrometer operates within the spectral range of $\SI{250}{\nano\metre}$ to $\SI{11.35}{\micro \metre}$, corresponding to $\num{5.5}$ optical octaves. This is achieved by dividing the signal beam into three spectral groups, each analyzed
by a dedicated spectrometer and detector unit. The complete instrument was characterized with regard to wavelength, relative spectral sensitivity, and absolute photo-metric sensitivity, always accounting for the light polarization and comparing different calibration methods. Finally, the capability of the spectrometer is demonstrated with a CTR measurement of a laser wakefield accelerated electron bunch, enabling to determine temporal pulse structures at unprecedented resolution.\\
\end{abstract}

%\pacs{Valid PACS appear here}% PACS, the Physics and Astronomy
                             % Classification Scheme.
%Use showkeys class option if keyword
                              %display desired
\maketitle
%\tableofcontents

\section{Introduction}\label{sec:Introduction}

Precise knowledge of the temporal structure of ultrashort electron bunches is key to understanding and control of modern linear accelerator driven light sources~\cite{Lumpkin2002} and high peak current applications like beam driven plasma wakefield accelerators~\cite{Chen1985,Joshi2003,Litos2014,Kurz2019}. On different energy scales it is essential for applied research such as ultrafast probing of transient processes in matter by ultrafast electron diffraction~\cite{Williamson1997,Jiang2020}. In particular the future refinement of laser wakefield accelerator (LWFA)~\cite{Tajima1979} and related light source concepts~\cite{Esarey2009,Corde2013,Downer2018} relies on improved diagnostic capability of the longitudinal phase space. It represents the main motivation of the current work as the temporal profile of wakefield accelerated bunches of a few femtosecond duration~\cite{Debus2010,Lundh2011,Buck2011,Heigoldt2015} with potential substructures~\cite{Luttikhof2010,Xu2016,Zarini2019} is closely linked to acceleration and injection conditions. Established bunch metrology techniques based on electro-optical sampling~\cite{Berden2004,Berden2007,Debus2010} or the use of streak cameras~\cite{Fesca100} are mostly limited to sub-100 fs temporal resolution, whereas a newly established method based on transverse deflecting structures (TDS) operated at X-band radio-frequency (RF) reached a resolution of better than 10 fs~\cite{Maxson2017,Marchetti2017}. However, the latter requires a non-trivial jitter-free bunch-to-RF synchronization as well as a relatively high RF-power. Both requirements represent a challenge for multi-GeV electron beams at very low repetition rate as typically generated in the LWFA regime~\cite{Gonsalves2019}. Though pulse-to-pulse stability in LWFAs has improved substantially in recent years, remaining fluctuations and in particular tracking changes require single-pulse availability of the complete information. Thus, in order to control beam quality and to improve the performance of plasma-based accelerators the development of single-shot, high-resolution spatio-temporal electron beam diagnostics is essential~\cite{Downer2018}.

The measurement of the spectral intensity of transition radiation (TR) emitted instantaneously when an electron bunch traverses a solid-vacuum interface~\cite{Ginzburg1946} is a promising method to deduce the longitudinal bunch characteristics on the sub-femtosecond time scale~\cite{Schmidt2018,Downer2018}. In addition to its single-shot capability, the temporal resolution of this technique does not depend on the beam energy. Hence it can be applied for MeV to GeV beams while the setup can be kept within a compact footprint. The TR spectrum contains information about the frequency dependent electron bunch form factor~\cite{Schroeder2004} from which the longitudinal charge distribution of the corresponding electron bunch can be reconstructed. The challenges linked to the approach lie in devising an absolutely calibrated multi-octave spectrometer with high spectral resolution and dynamic range, and in solving the non-trivial inverse problem of retrieving the temporal profile of the bunch from the measurement of spectral intensity alone. For the latter, advanced case-specific phase-retrieval algorithms for the reconstruction of the bunch have been developed~\cite{Bajlekov2013,Zarini2019}.

\begin{figure}[htbp]
\centering
\includegraphics[width=0.47\textwidth]{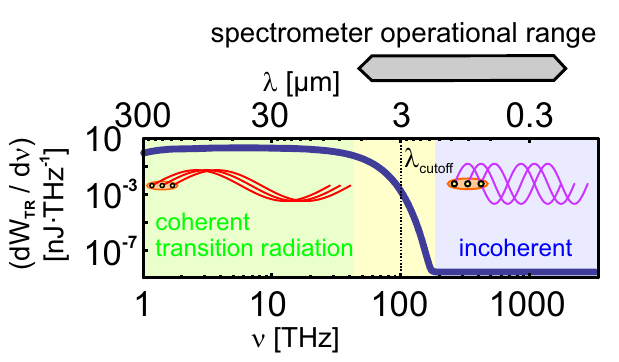}
\caption{Simulated spectral intensity of TR emitted by an electron bunch with \SI{20}{pC} charge, \SI{10}{fs} (FWHM) duration and \SI{200}{MeV} energy. Coherent transition radiation (CTR) is generated when the emitted wavelengths are longer than the bunch length, while incoherent transition radiation (ITR) dominates for shorter wavelengths. The  operational range of the spectrometer is indicated, covering $\SI{250}{\nano\metre}$ to $\SI{11.35}{\micro \metre}$. }
\label{fig:overview_em_spectrum}
\end{figure}

Here we present all steps of the realization of an ultra-broadband optical-spectrometer optimized to measure the TR spectrum over its full spectral beandwidth and with high resolution in single-pulse operation. This approach is essential for determining structural details and duration of an electron bunch in the range from $\SI{0.7}{fs}$ to $\SI{40}{fs}$. Fig.~\ref{fig:overview_em_spectrum} presents the simulated spectral-intensity of TR emitted by an electron bunch with a total charge of 20 pC in a smooth distribution of 10~fs full-width at half-maximum (FWHM) duration and 200~MeV energy. The corresponding spectrometer range is spanning from $\SI{250}{\nano\metre}$ to $\SI{11.35}{\micro \metre}$. The radiation features two distinct spectral regions, the coherent part (CTR) and the incoherent part (ITR), separated by a transition range centered at a characteristic wavelength $\lambda_\mathrm{cutoff}$~\cite{Schroeder2004} of $\SI{3}{\um}$. This wavelength corresponds to the bunch duration $\tau$ via $\tau\,\approx\,c\,\lambda_\mathrm{cutoff}$, $c$ being the speed of the light. The intensity of the CTR scales up to $\propto N^2$ while ITR is proportional to $N$, when $N$ represents the number of electrons within the bunch. Hence, for 20 pC of charge ($N\approx 10^8$ electrons) as in this example an intensity difference of up to 8 orders of magnitude is expected between the two regions. Measuring the absolute spectral-intensity over the full spectral range, i.e. the ITR, the transition, and the CTR ranges, is essential for a meaningful reconstruction of the bunch distribution. This thus requires a spectrometer equipped with sensitive detectors supporting high dynamic range over the full spectral bandwidth. Additionally, in order to resolve spectral modulations arising from potential bunch sub-structures at sub-fs time scales, it is mandatory to have high spectral resolution in parallel.

Due to the limited spectral bandwidth of existing detector technologies and the limited dispersion and transmission properties of optical materials, the use of a modular spectrometer is indispensable. In the presented instrument, the input spectrum is divided into middle infrared (MIR), near infrared (NIR), as well as visible (VIS) to ultraviolet (UV) fractions by utilizing carefully selected beam splitters. Each fraction is thereafter spectrally decomposed using common yet range adapted dispersive elements such as prisms and gratings. The partial spectra are recorded by means of three detectors and, after post-processing, recombined into the extended final spectrum.

To account for any variation in the spectral response of individual detectors, as well as for transmission and reflection properties of all optical elements a full characterization of the instrument is performed over the entire extended spectral range. This includes a wavelength calibration~\cite{Lerner2004,Martinsen2008,Du2011}, a relative spectral response calibration~\cite{Dorazio1974,Wilbur2007}, and finally an absolute photo-metric calibration. The latter is the essential step for calculating the electron bunch form factor in terms of the absolute intensity of the TR for the electron bunch duration measurements.

The paper is organized as follows. In Section~\ref{sec:Design_setup} design and setup of the spectrometer as well as a detailed review of spectrometer components are presented. In Section~\ref{sec:calibration} the full characterization of the spectrometer is described, presenting the three calibration steps. In Section~\ref{sec:exp_results} experimental results of TR measurements in LWFA experiments are shown, and summarized in Section~\ref{sec:conclusion}.

\section{Design and setup}\label{sec:Design_setup}

\begin{figure}[htbp]
\centering
\includegraphics[width=0.45\textwidth]{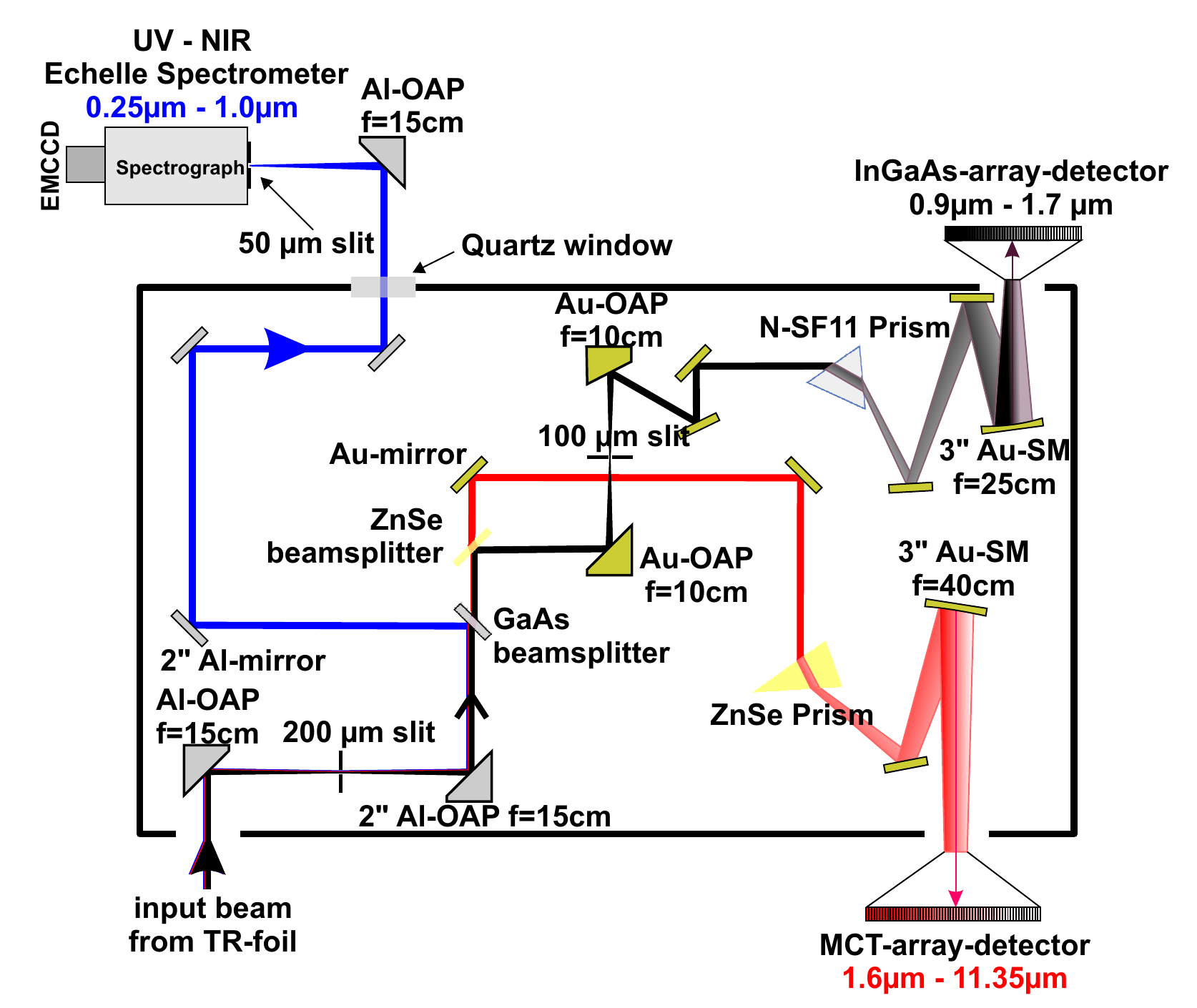}
\caption{Overview of the main components of the TR spectrometer showing a schematic top-view of the spectrometer design and highlighting the three main light paths. The vacuum chamber dimensions are \SI{120}{cm} x \SI{68}{cm} x \SI{40}{cm}. Additional optical elements such as polarizers, attenuators and bandpass filters, as well as imaging optics for monitoring the TR beam pointing used during the calibration and TR measurement are not shown here for clarity. Elements that have to be aligned during calibration or experimental runs are motorized.}
\label{fig:spectrometeroverview}
\end{figure}

A schematic overview of the instrument with its three independent spectrometer arms and all central components is presented in Fig.~\ref{fig:spectrometeroverview}. All optical components are installed inside a dedicated vacuum chamber for two reasons. First, this approach eliminates any absorption of the radiation of interest in air~\cite{Demtroeder2008} due to vibrational-rotational transitions of molecules such as $\text{H}_{\text{2}} \text{O}$ or $\text{CO}_{\text{2}}$ which fall into the range of $\num{3}$ to $\SI{10}{\micro \metre}$. Consequently, NIR and MIR detectors are directly attached to this vessel.
The thickness of optical base plate and chamber walls have been optimized using the software~\textsc{Ansis} to minimize deformations and thus to preserve sensitive spectrometer alignment after pumpdown. The maximum relative deformation of the based-plate ($\SI{10}{cm}$ aluminum) and the walls have been simulated to amount to less than $\SI{0.1}{mm}$ and $\SI{0.3}{mm}$, respectively. Second, operation in vacuum enables the direct connection of the instrument to the accelerator vacuum, avoiding the use of any bandwidth-critical transmission element for TR beam transport.

In operation, a TR beam actively collimated close to its source enters the instrument and first passes a Keplerian telescope arrangement consisting of two $\SI{90}{\degree}$ off-axis-parabolic mirrors (OAPs) with equal effective focal length ($\text{EFL}=\SI{15}{cm}$). A slit is placed at the focal plane of the first OAP to set the overal spectral resolution of the spectrometer~\cite{Wilson1995} (see Section~\ref{par:spectral_resolution}). This OAP effectively images the TR source onto the slit. The recollimated TR beam is thereafter spectrally divided by two consecutive beam splitters (see Section~\ref{sec:beamsplitters}). The first beam splitter, made of GaAs, reflects the UV to NIR region of the full TR spectrum, i.e. from $\SI{0.2}{\micro \metre}$ to $\SI{1.0}{\micro \metre}$, which is analyzed by an Echelle spectrometer (see Section~\ref{sec:mechelle}). The part of the spectrum transmitted through the GaAs is further divided by a ZnSe beam splitter into its MIR and NIR components. The reflected NIR fraction ($\SI{0.9}{\micro \metre}$ to $\SI{1.7}{\micro \metre}$) is passed through a second slit placed at the focal plane of a second telescope arrangement consisting of two gold coated OAPs. The recollimated beam is spectrally analyzed by a glass prism (N-SF11, see Section~\ref{sec:prisms}). The light is thereafter focused by a spherical mirror (SM) onto an InGaAs array detector (see Section~\ref{sec:InGaAs}). The remaining MIR fraction ($\SI{1.6}{\micro \metre}$ to $\SI{12}{\micro \metre}$) is directly sent to a ZnSe prism and focused by a SM onto a mercury cadmium telluride (MCT) array detector (see Section~\ref{sec:MCT}).

In the following, individual selection criteria for the components of the three spectrometer arms are discussed.

\subsection{Basics of a prism spectrometer}\label{sec:basics_of_prism_spec}
\label{par:spectral_resolution}

\begin{figure}[htbp]
\centering
\includegraphics[width=0.48\textwidth]{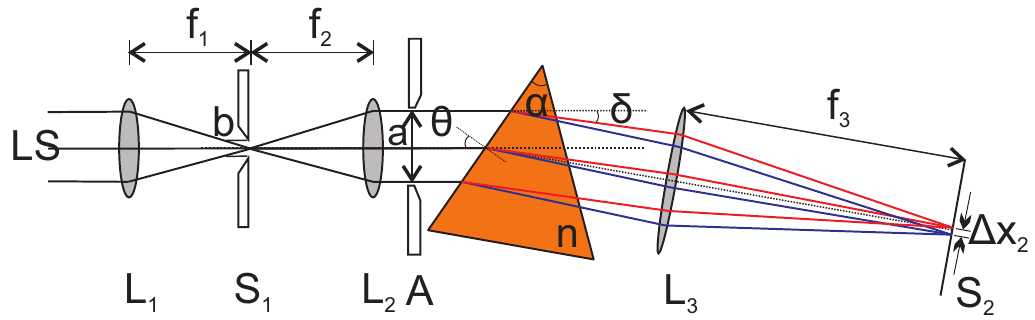}
\caption{Schematic of a prism spectrometer.}
\label{fig:basics_prism_spectrometer}
\end{figure}

The concept of a prism spectrometer is depicted in Fig.~\ref{fig:basics_prism_spectrometer}. A collimated light source $\text{LS}$ is focused by lens $\text{L}_1$ and illuminates an entrance slit $\text{S}_1$, placed in the focal plane (focal length $\text{f}_1$) of $\text{L}_1$. The beam is recollimated by $\text{L}_2$ and passes through the prism $\text{P}$, where it is refracted by an angle $\delta(\lambda)$ depending on the wavelength $\lambda$, the incidence angle on the prism $\theta_1$ and its apex angle $\alpha$. The camera lens $\text{L}_3$ images the entrance slit $\text{S}_1$ onto the detector array $\text{S}_2$, providing the wavelength dependent position signal $x(\lambda)$. The linear dispersion $dx/d\lambda$ of the spectrometer depends on the spectral dispersion $dn/d\lambda$ of the prism material and on the focal length $f_3$. Depending on the size of the detector array $\Delta x = x_1 - x_2$, a spectral range $\Delta \lambda = \lambda _1 (x_1)- \lambda_2 (x_2)$ can be covered.

In the prism the wavelength pair $\lambda$ and $\lambda+\Delta\lambda$ is split into beamlets with angular deviations $\delta$ and $\delta +\Delta\delta$. The angular separation is defined by the angular dispersion $d\delta/d\lambda$ (in units of [$\si{rad/nm}$]) as
\begin{equation}
\Delta \delta = \left( \dfrac{d\delta}{d\lambda} \right)\Delta\lambda .
\end{equation}

The camera lens then images the entrance slit $\text{S}_1$ onto the plane $\text{S}_2$. The distance $\Delta x_2$ between the two images $S_2(\lambda)$ and $S_2(\lambda+\Delta \lambda)$ is thus obtained by
\begin{equation}\label{eq:deltax2}
\Delta x_2 = f_3 \Delta \delta = f_3 \dfrac{d\delta}{d \lambda} \Delta \lambda = \dfrac{dx}{d \lambda} \Delta \lambda ,
\end{equation}
where $dx/d\lambda$ denotes the linear dispersion of the spectrometer (in units of [$\si{\mm}/\si{\nm}$]).

\begin{figure}[htbp]
\centering
\includegraphics[width=0.45\textwidth]{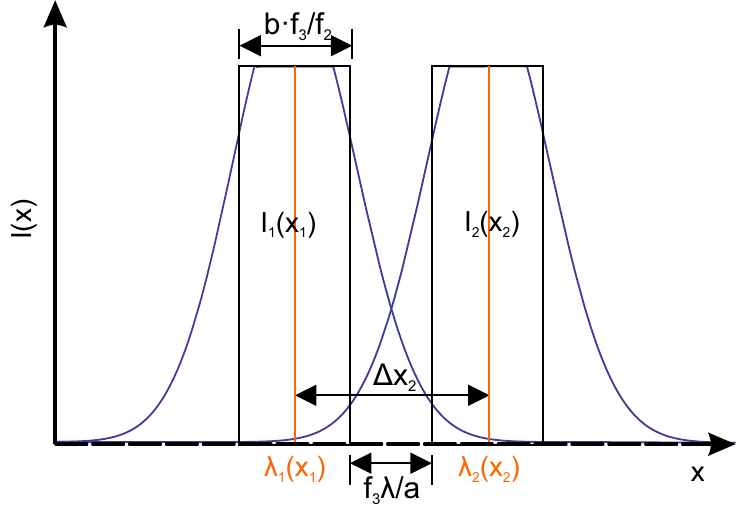}
\caption{Illustration of the resolution limit of two closely overlapping spectral lines, where $x$ denotes a position on the spectrometer detector and corresponds to some wavelength $\lambda(x)$.}
\label{fig:2gaussian_resolution}
\end{figure}

The spectral resolution is assumed to be dominated by diffraction depending on the aperture $a$ and focal length $f_3$ (see Fig.~\ref{fig:2gaussian_resolution}). According to the Rayleigh criterion \cite{Hutley1982} the minimum distance between two resolvable slit images reads as \cite{Demtroeder2008}
\begin{equation}\label{eq:difflimit}
\Delta x_R = f_3 \left( \lambda / a \right).
\end{equation}
Further taking into account the finite size of the slit $b$ and applying geometrical optics, the width of the image of the slit is obtained by
\begin{equation}\label{eq:geometrylimit}
\Delta x_b= b \left( f_3/f_2 \right).
\end{equation}

The separation $\Delta x_2$ between the peaks of the two image distributions $I_1(\lambda_1)$ and $I_2(\lambda_2)$ must this exceed the sum of the above (see Fig.~\ref{fig:2gaussian_resolution})
\begin{equation}
\Delta x_2 \geq f_3 \dfrac{\lambda}{a}+ b \dfrac{f_3}{f_2}.
\label{eq:minseparation}
\end{equation}
The smallest resolvable wavelength interval $\Delta \lambda$, the spectral resolution, is obtained from Eq.~\eqref{eq:deltax2} as
\begin{equation}
 \Delta \lambda \geq \left( \dfrac{\lambda}{a}+\frac{b}{f_2} \right) \left(\dfrac{d \delta}{d \lambda} \right)^{-1}.
\label{eq:spectral_resulution}
\end{equation}
Note, that for an infinitely small entrance slit $b$ the spectral resolution is dominated by the diffraction caused by the much larger aperture $a$ (Eq.~\eqref{eq:difflimit}), typically given by the size of the beam optics. Thus, optimum slit width settings can be chosen to be compliant with transmission efficieny optimization of the apparatus.

Finally, the spectral resolving power $R$ is derived from the spectral resolution to
\begin{align}
R =  \bigg|\dfrac{\lambda}{\Delta \lambda} \bigg| &= \lambda \left(\dfrac{\lambda}{a}+\dfrac{b}{f_2} \right)^{-1} \left(\dfrac{d\delta}{d\lambda} \right) \nonumber \\
                              &= \lambda \left( \dfrac{\lambda}{a}+\dfrac{b}{f_2} \right)^{-1} f_3^{-1} \left( \dfrac{dx}{d\lambda} \right).
\label{eq:resolving_power_linear}
\end{align}

\subsection{Beam splitters} \label{sec:beamsplitters}
Dichroic beam splitters are normally designed for a specific spectral bandwidth either in the VIS or IR range and thus not applicable for the separation of broad spectra. By using uncoated infrared transparent bulk material and taking advantage of its Fresnel reflection and transmission properties, two consecutive beam splitters are selected in order to spectrally separate the input beam into three spectral bands (see Fig.~\ref{fig:spectrometeroverview}).
For the first beam splitter, a $\SI{6}{\mm}$ thick Gallium-Arsenide-plate (GaAs) with $\num{12}$ minutes wedge angle (manufacturer \textrm{II}-\textrm{VI}-\textsc{Infrared}) is used to separate the UV-NIR range of the spectrum ($\SI{250}{\nm}$ to $\SI{1.0}{\micro\meter}$) from the TR beam. The small remaining fraction of Fresnel reflected longer wavelength light is further attenuated by the fused silica exit window (Corning 7980).

For the NIR and MIR Zinc-Selenide (ZnSe)~\cite{Marple1964} is the preferred material for transmision optics because of its low absorptivity at infrared wavelengths ($\leq \SI{0.0005}{\cm^{-1}} @ \SI{10.6}{\micro \metre}$) and its acceptable visible light transmission for alignment purposes. For the second beam splitter, a $\SI{3}{\mm}$ thick ZnSe plate with $\num{6}$ minutes wedge angle (manufacturer \textsc{Korth Kristalle}) is utilized.

\begin{figure}[htbp]
\centering
\includegraphics[width=0.5\textwidth]{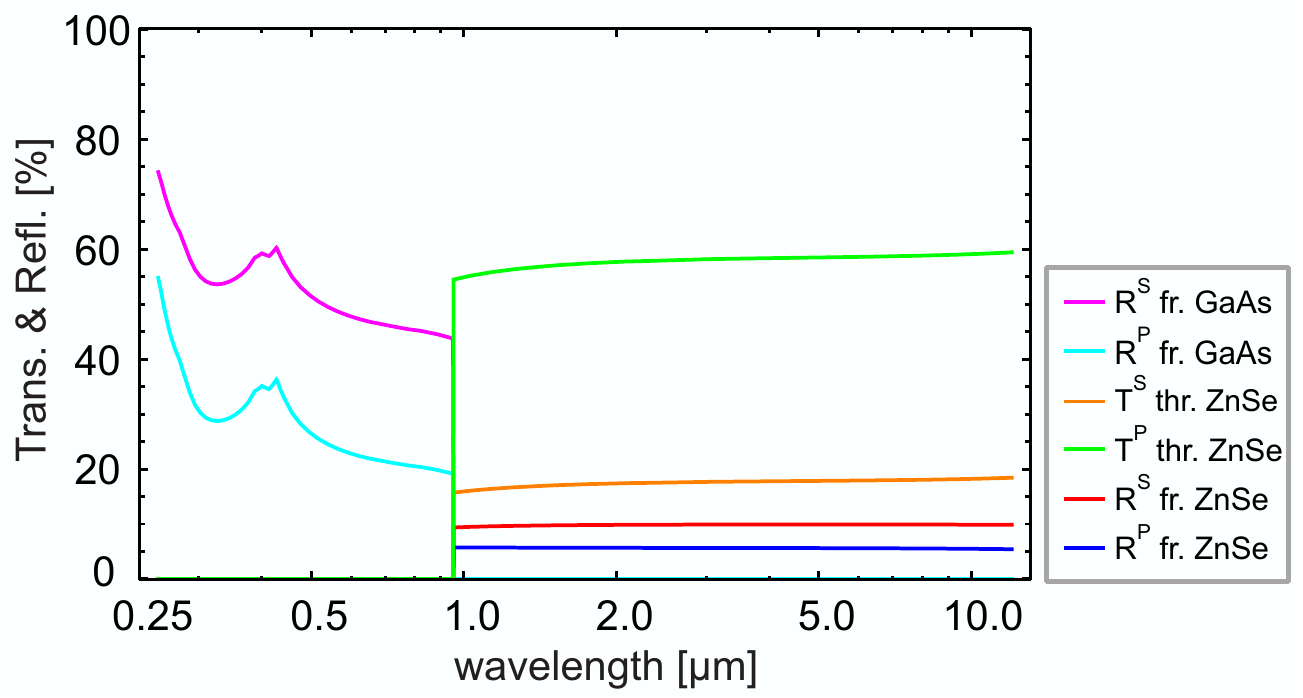}
\caption{Input beam transmission and reflection for the GaAs and ZnSe beam splitters. GaAs, with a sharp transmission edge at around $\SI{900}{\nm}$ and low bulk absorption of $\leq \SI{0.01}{\cm^{-1}}$ @$\SI{10.6}{\um}$, is ideal for a spectral separation of the UV-NIR range from the near to the MIR range. ZnSe reflects on average $\SI{7}{\%}$ of the beam and transmits $\SI{18}{\%}$ and $\SI{60}{\%}$ of the input beam for s- and p-polarization, respectively.}
\label{fig:BS_transmission}
\end{figure}

Transmission and reflection parameters are calculated by means of Fresnel equations \cite{Meschede2017} with respect to s- and p-polarization (in the spectrometer plane). The wavelength dependent refractive indices of these materials are modeled applying Sellmeier-polynomials \cite{Ghosh1997}. The absorption as well as losses from secondary internal reflections in the GaAs material is taken into account while absorption in ZnSe bulk material is negligible. The results of this calculation are shown in Fig.~\ref{fig:BS_transmission}.

\subsection{Prism optimization} \label{sec:prisms}
The free spectral range of a spectrometer $F_{\lambda}$ is defined as the wavelength range for which an unambiguous relationship between $\lambda$ and the position $x(\lambda)$ at the detector surface exists. A single prism based instrument covers a broad spectral range depending on its dispersion $n(\lambda)$ at a resolution
that is rather low compared to grating based solutions. For the latter, $F_{\lambda}$ is in contrast limited by overlapping diffraction orders. Furthermore, prisms
exhibit higher throughput and lower stray light characteristics. For customizing the geometry of a prism and selecting its material, the following optimization criteria are applied:

\textbf{1.} The free spectral range $F_{\lambda}=\lambda_\text{high}-\lambda_\text{low}$ is set with respect to the spectral response of available detectors, including transmission and reflection properties of the optics.

Due to a good quantum efficiency (QE) of the InGaAs detector (see Section~\ref{sec:InGaAs}) in the range of $\SI{800}{\nm}$ to $\SI{1.7}{\micro \metre}$ (limits @ $\SI{20}{\%}$ level), and the cutoff of the GaAs beam splitter at $\SI{900}{\nm}$ (see Section~\ref{sec:beamsplitters}) the spectral range for the NIR arm is set from $\SI{900}{\nm}$ to $\SI{1.7}{\micro\metre}$.

According to the spectral response of the MCT detector (see Section \ref{sec:MCT}) the spectral range of $\SI{1.6}{\micro\metre}$ to $\SI{11.35}{\micro\metre}$ is defined for the MIR arm including some spectral overlap of $\SI{100}{\nm}$ between NIR and MIR ranges, facilitating the later absolute cross-calibration of the spectrometer arms.

\textbf{2.} The linear dispersion $dx/d\lambda$ introduced by the prism should cover the entire detector array with regard to the focal length of the focusing optic. Here spherical mirrors (SM) are utilized in order to minimize imaging errors across the spectrum. As the SM is irradiated at a small off-axis angle (in our case $\sim \SI{10}{\degree}$) the choice causes a slight astigmatism leading to elliptical foci on the detector array plane. As these foci are smaller than the size of the InGaAs pixels ($500\times25$ \si{\um}) no intensity loss is expected.

\textbf{3.} The prism should feature linearity in dispersion for achieving a uniform spectral resolution on the corresponding detector array. The non-linearity (NL) of the dispersion is defined by the spectral sampling ratio (SSR)
\begin{equation}
\mathrm{SSR}=\mathrm{max}\left\vert \frac{\mathrm{d}\delta}{\mathrm{d}\lambda}\right\vert / \mathrm{min}\left\vert \frac{\mathrm{d}\delta}{\mathrm{d}\lambda}\right\vert,
\end{equation}
representing a good figure of merit for the optimization of the prism geometry.

\begin{figure}[htbp]
\centering
\includegraphics[width=0.4\textwidth]{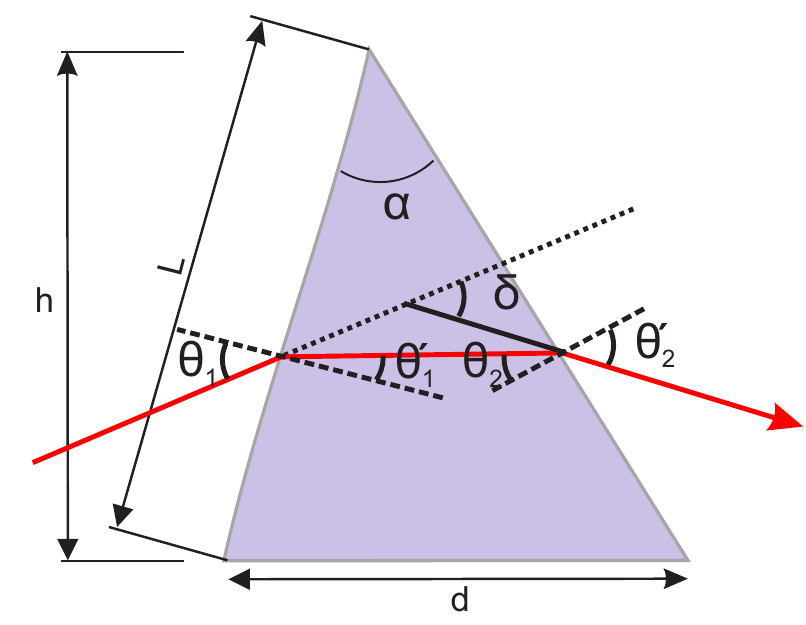}
\caption{Ray trace through a prism. $\theta_1, \theta_1^{\prime}, \theta_2, \theta_2^{\prime}$ are the incident and refractive angles at the first and second prism interface, respectively, $\alpha$ the prism apex angle, $d, L, h$ are the geometric dimensions of the prism, and $\delta$ is the deflection angle with respect to the incident beam. Note the sign convention of the angles $\theta$ and $\delta$, in which clockwise angles with respect to the surface normal have negative sign and vice versa.}\label{fig:prismconfig}
\end{figure}

\textbf{4.} Large angles ($\theta_1$ and $\theta_2^{\prime}$ in Fig.~\ref{fig:prismconfig}) on the prism cause large projected beam diameters requiring large prism optics, as well as precise angular alignment. Additionally, these scenarios lead to significant losses due to reflection on the prism surfaces. For excluding such grazing incidence designs the constraints $\theta_1 < \SI{60}{\degree}$ and $\theta_2^{\prime} < \SI{60}{\degree}$ are assumed for prism optimization.

\textbf{5.} Finally, the prism geometry needs to be chosen such that its refractive surfaces are compliant with the beam diameter and the overall size remains reasonable.

\begin{figure*}[htbp]
\centering
\includegraphics[width=0.95\textwidth]{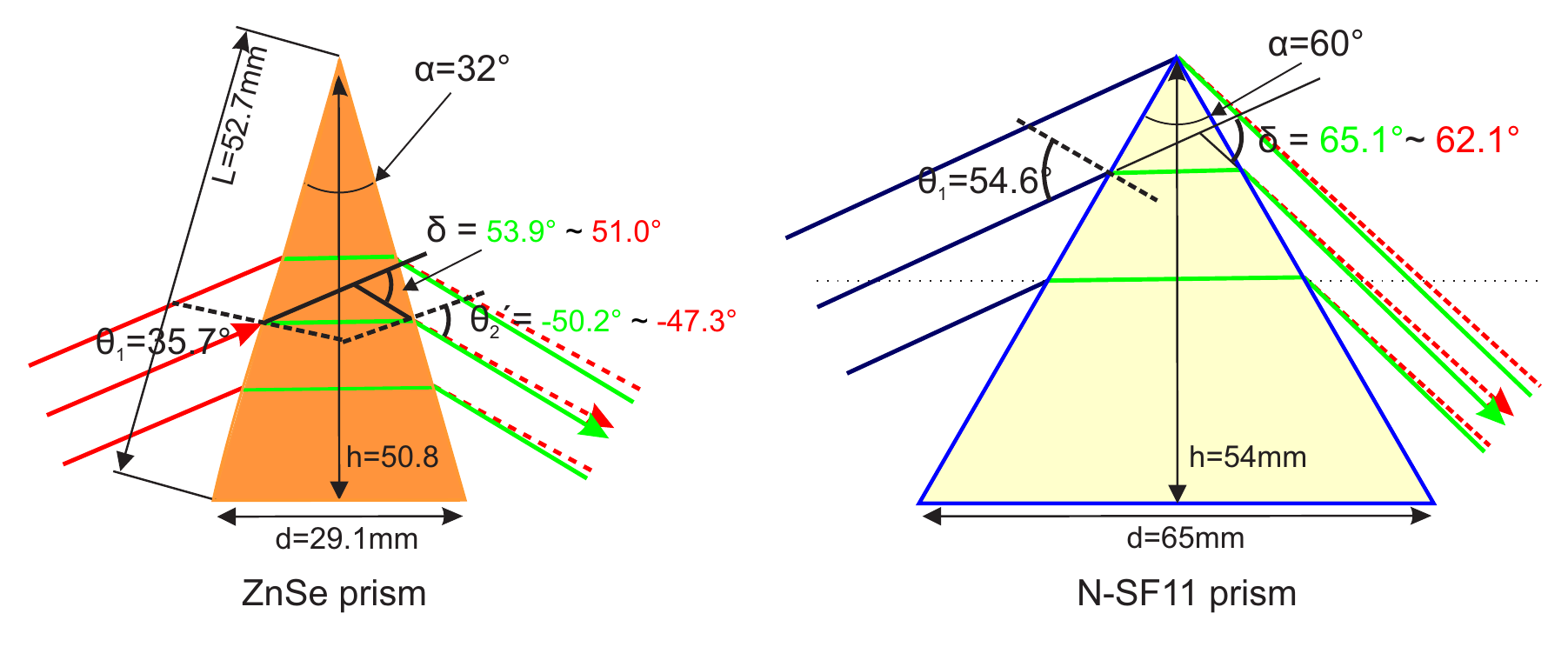}
\caption{Illustration of N-SF11 and ZnSe prism geometries, designed for NIR ($\SI{0.9}{\um}$ to $\SI{1.7}{\um}$) and MIR ($\SI{1.6}{\um}$ to $\SI{12}{\um}$) ranges of the spectrometer, respectively.}
\label{fig:prismdesign}
\end{figure*}

The prism geometry is modeled in \textsc{Mathematica}~\cite{Mathematica2009} following geometrical optics (see Fig.~\ref{fig:prismconfig}) and ray paths are calculated with the help of Snell's law.
The deflection angle of a beam behind the prism can be expressed as \cite{hagen2011compoundI}
\begin{eqnarray}\label{eq:prism_design_deflection}
\delta \left( \lambda,\theta_1, \alpha \right) &= &- \alpha+ \theta_1 \\
                                               &  &+\arcsin\left( n(\lambda) \sin\left( \alpha - \arcsin\left( \dfrac{\sin (\theta_1)}{n(\lambda)} \right) \right) \right). \nonumber
\end{eqnarray}
A comprehensive description of the optimization criteria can be found in Refs. \cite{hagen2011compoundI, hagen2011compoundII, hagen2011compoundIII}. A global optimization algorithm based on the differential evolution method was employed in order to determine the prism geometry for a series of available infrared materials. The selected prism designs, N-SF11~\cite{NSF11index} for the NIR range and ZnSe~\cite{MarpleZnSeindex} for the MIR range, are documented in Fig.~\ref{fig:prismdesign}. In principle, the spectral linearity of the prism spectrometer could be improved by a double prism design, in particular by the combination of ZnSe and NaCl prisms. As it turned out the remaining gap between the prisms, due to the lack of an interfacing glue material suitable for the required spectral range, leads to multiple reflections. Furthermore the hygroscopic property of NaCl is disadvantageous. Thus this option was discarded for the MIR arm.

\subsection{Echelle spectrometer} \label{sec:mechelle}
An Echelle spectrometer supports unprecedented bandwidth in single pulse operation with high resolution and spectral sensitivity by combining a prism with a grating subsequently acting in orthogonal dispersion planes. The echelle grating diffracts the spectrum into several orders, each of which represents a portion of the high resolution spectrum. The prism then enables order separation when imaged onto a 2D detector array.
A commercial Mechelle spectrometer ME5000~\cite{MechelleManual} (\textsc{Andor Technology}) is implemented in the TR spectrometer for the analysis of the UV-NIR range ($\SI{200}{\nm}$ to $\SI{1030}{\nm}$). The input beam is focused through a $\SI{50}{\micro\metre}$ slit at the entrance of the echelle spectrograph, then recollimated by a F/7 SM and subsequently refracted by a double prism. The dispersed beam in turn falls onto a grating and is thereby diffracted perpendicularly to the refraction plane of the prism. A focusing mirror images the entrance slit onto a 2D EMCCD detector (see Section~\ref{ph:EMCCD}). The grating has a groove density of $\SI{52.13}{\mm^{-1}}$ and is used at high grating orders, i.e. $\num{20}$ to $\num{100}$ and blazed under $\SI{32.35}{\degree}$.

\begin{figure}[htbp]
\centering
\includegraphics[width=0.5\textwidth]{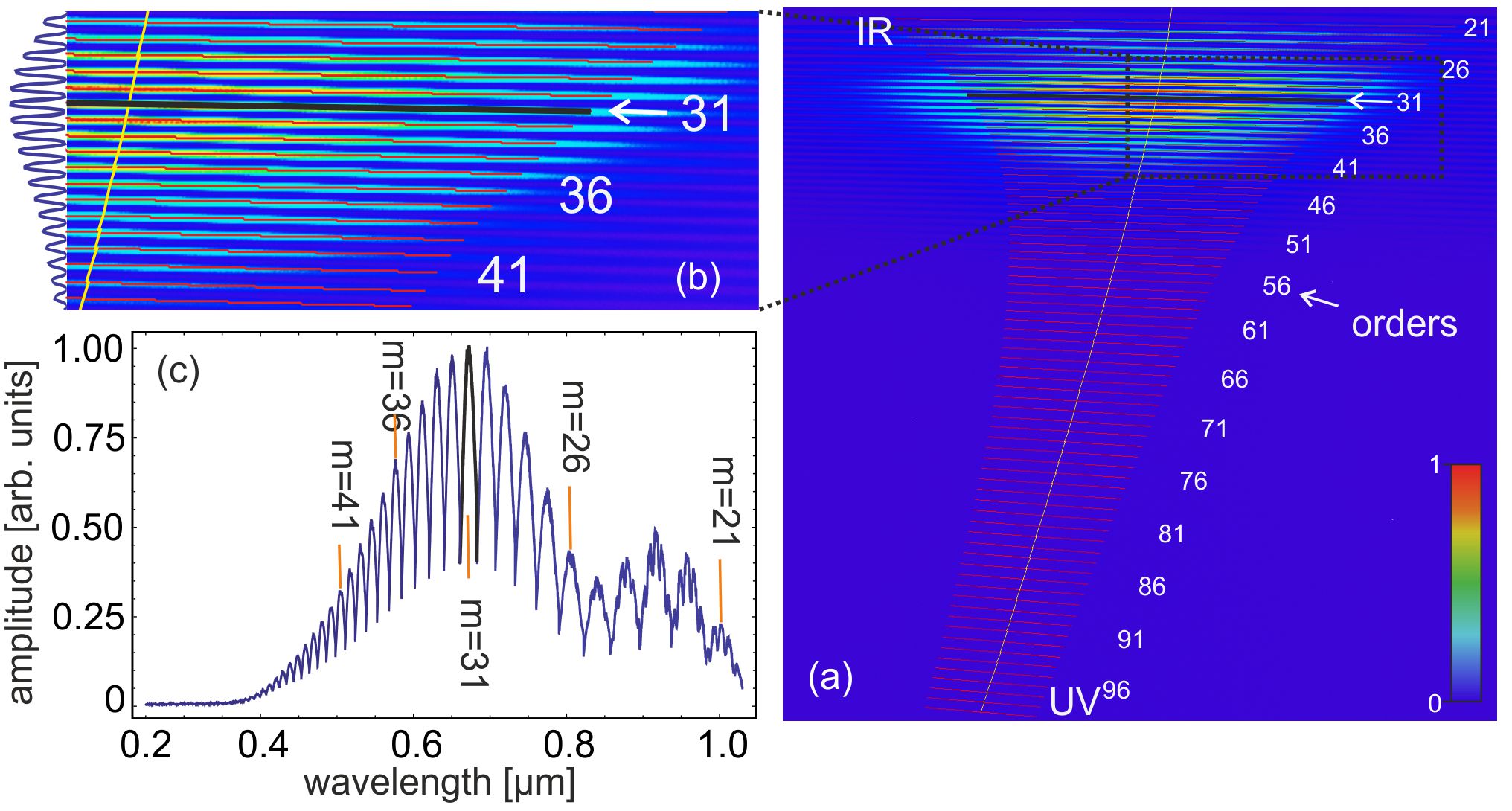}
\caption{Sample measurement of the UV-VIS spectrum using a QTH lamp. (a) illustrates the full image acquired with the 2D detector of the echelle spectrometer. The horizontal red lines indicate the (numbered) diffraction orders. The orange line connects the central wavelength $\lambda_m^0$ of each order. (b) shows a zoomed part of (a) including a vertical line out showing the separation of the diffraction orders. The vertical extension of individual orders covers around $\num{10}$ pixels. (c) presents a complete  spectrum sampled from all orders as detailed in section~\ref{sec:mechelle} and \ref{subsec:mechelle}.}
\label{fig:mechelle_2Dorderplots}
\end{figure}

The free spectral range for a single order $m$ is given by
\begin{equation}\label{eq:mechelle_order_BW}
F^m_{\lambda}=\dfrac{\lambda_m^{o}}{m \left( 1- (2m)^{-2} \right)},
\end{equation}
where $\lambda_m^{0}$ denotes the central wavelength of this order.
The overall bandwidth thus is the sum of the number of orders that can be covered by the 2D detector chip. The efficiency of the echelle spectrometer depends mainly on the efficiency of its grating and the QE of the coupled detector.  Although its grating deflection angle $\phi$ is not varying much over a certain spectral order ($\leqslant \SI{2}{\degree}$) \cite{Gaigalas2009,Bibinov2007}, the reflection coefficient of the grating and therefore the grating efficiency considerably drops from its maximum in the middle of each order towards the edges of each spectral order to about half of its value (see Fig.~\ref{fig:mechelle_2Dorderplots} (c)).
The strong variations in the spectral efficiency of the echelle spectrometer over its entire spectral range is illustrated in Fig.~\ref{fig:mechelle_2Dorderplots}, where a Quartz-Tungsten-Halogen (QTH) lamp is used as broad calibration source.

In the 2D CCD raw data, each spectral order approximately follows an approximately straight line with some fixed width. Thus for data analysis, we extract the center line of maximum intensity for each order and sum up the signal over the line width. At this stage, the resulting 1D signal arrays of all orders are still spectrally overlapping and have no well defined end-points. After wavelength calibration using a spectral line source, the spectral endpoints of each order are determined by wavelength position, at which either the pre- or proceeding order becomes spectrally more sensitive. All these such wavelength calibrated and truncated order-spectra are then joined to a single spectrum, see Fig.~\ref{fig:mechelle_2Dorderplots}(c).

\subsection{Detectors} \label{sec:detectors}

In order to record information over the broad spectral range of the instrument at high resolution and high sensitivity simultaneously we utilize an MCT array detector (for $\SI{1.6}{\micro\metre}$ to $\SI{11.35}{\micro\metre}$), an InGaAs array detector ($\SI{0.9}{\micro\metre}$ to $\SI{1.7}{\micro\metre}$), and an EMCCD detector ($\SI{250}{\nano\metre}$ to $\SI{1.0}{\micro\metre}$) as detailed in the following.

\subsubsection{EMCCD detector}\label{ph:EMCCD}
The 2D spectrum of the echelle spectrograph (see Section~\ref{sec:mechelle} ) is recorded with a $\num{14}$-bit, back illuminated Electron Multiplying Charged Coupled Devices (EMCCD detector, iXon3 888, model A-DU888-EC-BVF, \textsc{Andor Technology}). Its active area consists of $1024\times1024$ pixels with a pixel size of $13\times13\,\si{\um}$.
The EMCCD sensor technology enables the measurement of very weak signals by suppressing the readout noise during the readout process. This is a trade-off reducing the camera's effective dynamic range as the feature is realized by amplifying the signal in electron multiplier (EM) registers before it is transferred to the output amplifier and A/D-converter. By suppressing the readout noise using an EM gain of 30, an effective dynamic range larger than 14~bit could be maintained.
The detector head is enclosed in a vacuum container behind an AR coated fused silica window. Furthermore a built-in thermocouple cools the chip down to $\SI{-95}{\celsius}$ reducing the (thermal) dark current in the detector by more than one order of magnitude.

\subsubsection{Indium-Gallium-Arsenide (InGaAs) detector} \label{sec:InGaAs}
For the NIR detection in the range of $\SI{0.9}{\micro\metre}$ to $\SI{1.7}{\micro\metre}$ with high QE, minimized noise, and a high dynamic bandwidth, an InGaAs array-detector (\textsc{Andor Technology}, iDus $\num{490}$) is used. The detector array consists of 512 pixels, each $\SI{25}{\micro\metre}$ wide and $\SI{500}{\micro\metre}$ high. As the EMCCD the detector array is mounted in a dedicated vacuum invironment, cooled to $\SI{-90}{\celsius}$, and accessed through an uncoated fused silica window.

\subsubsection{Mercury-Cadmium-Telluride (MCT) detector}\label{sec:MCT}
In order to measure the spectrum in the MIR range from $\SI{1.6}{\micro\metre}$ to $\SI{11.35}{\micro\metre}$, a photoconductive (PC) MCT array detector (\textsc{Infrared Systems}, IR-6416) is applied. The semiconductor-based MCT detector is more sensitive and faster than thermal (pyroelectric) detectors featuring an orders of magnitudes higher peak specific detectivity $D^\star=\num{2e+10}\,\text{cm}\,\text{Hz}^{1/2}/\,\text{W}$ \cite{Rogalski2010,Budzier2011}.
Peak response was customized by the manufacturer from $\SI{2}{\micro\meter}$ to $\SI{14}{\micro\metre}$ (@ $\SI{20}{\percent}$ spectral response) by adjusting the alloy composition of its ternary compound. The MCT detector array consists of 64 pixels. The active area of each pixel is $\SI{300}{\micro\metre}$ wide and $\SI{600}{\micro\metre}$ high.  An additional inactive space of $\SI{25}{\micro\metre}$ separates the pixels from each other.
The chip is housed in a Dewar flask behind an AR-coated ZnSe window and cooled with liquid nitrogen $\mathrm{LN_2}$ to $\SI{74}{\kelvin}$. The integration time window for recording can be set in $\SI{10}{\ns}$ steps from $\SI{54}{\ns}$ to $\SI{2614}{\ns}$. It should be noted that the MCT electronic measures changes in resistance. The readout amplifier thus uses as reference the average bias current from the $\SI{30}{\micro \s}$ preceding the integration time window for data acquisition. This measurement method is well suited for signals which occur within a $\si{ns}$-$\si{\micro \s}$ integration time. However, in order to perform a decent measurement regarding the absolute energy of a signal, the incidence pulse must be shorter than the time constant of the detector, i.e. $< \SI{400}{\ns}$.

\section{Calibration} \label{sec:calibration}

Providing an absolute calibration of spectral sensitivity of the full instrument, in units of measured detector signal per energy per spectral bandwidth $[\si{\text{(counts or V)}\times\um\per\joule}]$, is complex and divided into three separate tasks.

In the following, the calibration refers to all radiation passing through the \SI{200}{\um} input slit of the spectrometer within its acceptance half angle of the collection optics, which is \SI{16}{mrad} in the setup presented here, see Fig.~\ref{fig:spectrometeroverview}.

\begin{figure*}[htbp]
\centering
\includegraphics[width=.95\textwidth]{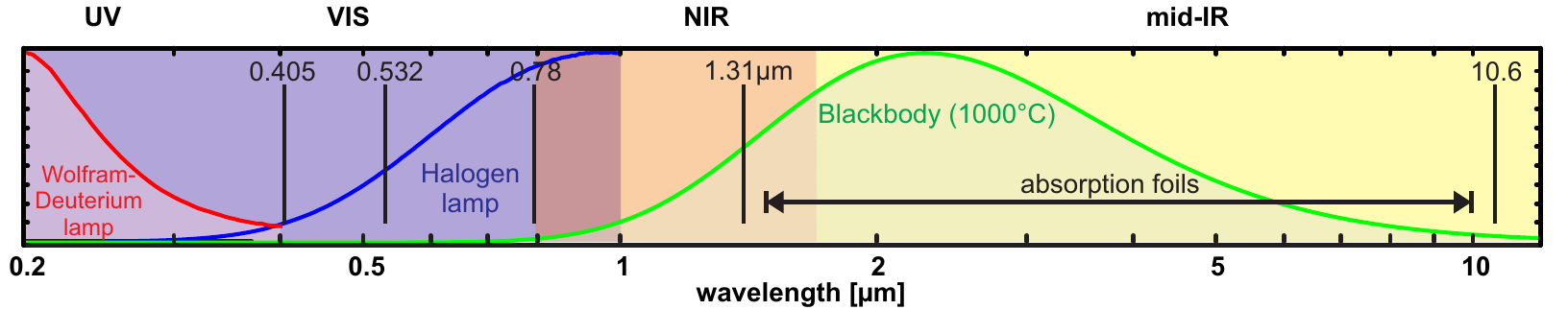}
\caption{Overview of the light sources used for the different steps of the calibration procedure comprising continuous light sources (LSs), laser lines (\textit{black lines}) and a set of absorption foils (TPX, PP, HDPE and Mylar). The spectra of the HgAr and Ar spectral lamps are not shown here.}
\label{fig:calibration sources}
\end{figure*}

The complex procedure providing absolute calibration of spectral sensitivity of the full instrument is divided into three separate tasks. Following wavelength calibration, a relative sensitivity calibration is performed. For selected wavelengths a final calibration based on photo-metric measurements transforms the relative calibration into an absolute. All steps include transmission and reflection properties of all optical elements as well as the spectral response and quantum efficiency of the individual detectors and distinguish between s- and p-polarization of the incoming light. Consequently, a wide variety of calibration sources is used. With an exception for the echelle spectrometer all calibration sources are coupled into the spectrometer through its entrance. An overview of the calibration sources is depicted in Fig.~\ref{fig:calibration sources}. These include monochromatic laser sources, calibrated spectral emission lines sources, characterized absorption foils and bandpass filters (BPF), as well as calibrated continuous sources such as black body, Quartz-Tungsten-Halogen (QTH) and Deuterium lamps. The wavelength calibration in the MIR range is performed by means of absorption lines of plastic films ($\sim\SI{10}{\um}$ thick) back illuminated by a black body (BB) reference spectrum. Spectral emission from an Argon source (Ar) and a Mercury-Argon source (HgAr) is used in the NIR and UV-VIS ranges, respectively. For the relative calibration photometrically defined broadband sources are employed, yet, in order to minimize errors, only the shape of the respective spectrum is exploited and not the photo-metric calibration. The absolute calibration is instead performed via a number of different continuous-wave (CW) lasers at selected wavelengths.

The wavelength calibration is discussed in Section~\ref{sec:wl_calibration} where the results are subsequently presented for each spectrometer arm. The relative as well as absolute calibration procedures follow in Sections~\ref{sec:rel_calibration} and \ref{sec:abs_calibration}.

The goal of the calibration is to provide the spectral sensitivity of the full instrument in units of measured detector signal per energy per spectral bandwidth $[\si{\text{(counts or V)}\times\um\per\joule}]$ and, as well as the spectral noise-equivalent energy in units of energy per spectral bandwidth $[\si{\joule/\um}]$ for estimating signal to noise ratios and minimal detectable signals.
Before IR calibration measurements are started, the TR spectrometer chamber is purged by dried nitrogen over several hours. This removes, in particular, moisture and creates a much drier calibration environment and minimizes IR absorption.

\subsection{Wavelength calibration} \label{sec:wl_calibration}
The wavelength calibration of a spectrometer involves the measurement of spectral lines at well-known wavelengths from the emission of calibration light sources (LS)~\cite{Martinsen2008} or absorption lines of a known absorber material. The relationship between the wavelength and the corresponding pixel positions on the detector is obtained by a polynomial fit~\cite{Du2011}. In the following we discuss a specific protocol for each of the spectrometer arms separately.

\subsubsection{Echelle spectrometer}\label{subsec:mechelle}
\begin{figure}[htbp]
\centering
\includegraphics[width=0.49\textwidth]{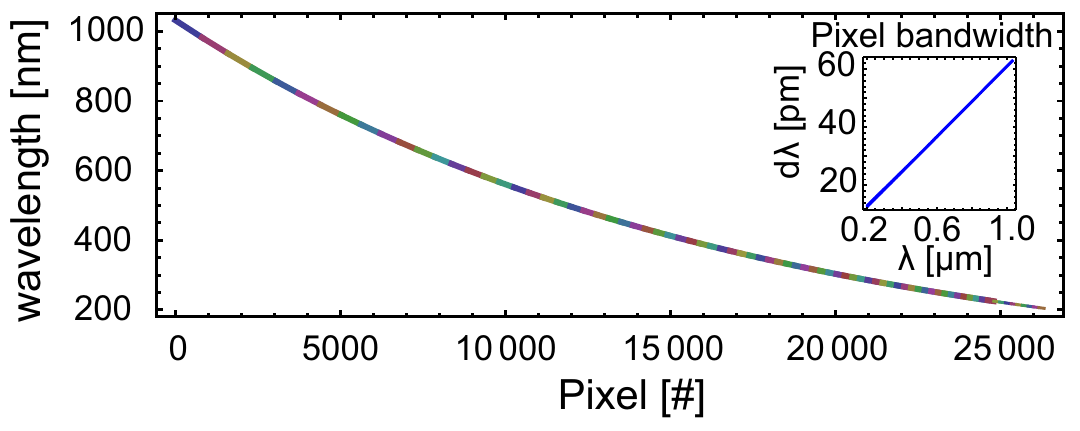}
\caption[Wavelength calibration of Echelle]{Wavelength calibration fit for the echelle spectrometer: Each color-distinguished partition of the curve represents a diffraction order. The inset here and in the following figures shows the spectral bandwidth covered by individual pixels as a function of wavelength.}
\label{fig:Mechelle_WLcal}
\end{figure}

A HgAr LS emits first order mercury and argon lines ranging from $\SI{253}\nm$ to $\SI{922}{\nm}$ and second order argon lines up to $\SI{1700}{\nm}$. A $\SI{50}{\micro\metre}$ diameter fiber optic (Andor, A-ME-OPT-8004) is used to couple the source directly into the echelle spectrometer.  Identification of these spectral lines on the EMCCD-detector is performed by using the implemented algorithm for wavelength calibration in the acquisition software provided by \textsc{Andor Solis}. Following the search of the relevant spectral lines, the boundaries of each diffraction order with respect to their central wavelengths $\lambda_m^0$ are thereby determined. Since the spectral bandwidth of any order is small compared to its entire spectral range, linear interpolation on each order is performed to accomplish the wavelength calibration of individual diffraction orders. The overall calibration is then carried out by combining the diffraction orders and performing a $5^{\text{th}}$-order polynomial fit. The calibration curve addresses the central pixel for each wavelength from which $\pm \num{5}$ pixels in the corresponding column, perpendicular to the orders, have to be summed up in order to read out the total spectral intensity (see Section~\ref{sec:abs_calibration}). As presented in Fig.~\ref{fig:Mechelle_WLcal}, while the spectral bandwidth of each order is decreasing toward shorter wavelengths, the resolution is increasing with the order number.

Our data analysis of the CCD raw data generally follows a procedure similar to the manufacturers data acquisition software. However we improved the data extraction of the individual spectral orders, as the endpoints of each order-line were not always stable over time (temperature dependence), in practice drifting by as much as 3 pixels vertically. While this usually is a negligible error in wavelength, it is relevant for the relative and absolute photometric calibration described later, which requires to capture the full signal. Failing to account for this drift immediately generates distinct artifacts in the final calibrated signal, that show as characteristic variations of the echelle efficiency pattern being imprinted on the measured signal.

This is solved by fine-tuning the region of interests in data extraction, which for each order is defined by two endpoints determined during calibration and a line height of 5~pixels. For any measured dataset, we thus vary the vertical positions of each spectral endpoint by up to $\pm3\,$pixels and compare the spectrally and vertically integrated raw signals, optimizing for the configuration with maximum signal. These resulting corrections change only gradually over time and order number and essentially ensure that the full height of the respective spectral orders is extracted.

\subsubsection{InGaAs detector}
\begin{figure}[htbp]
\centering
\includegraphics[width=0.49\textwidth]{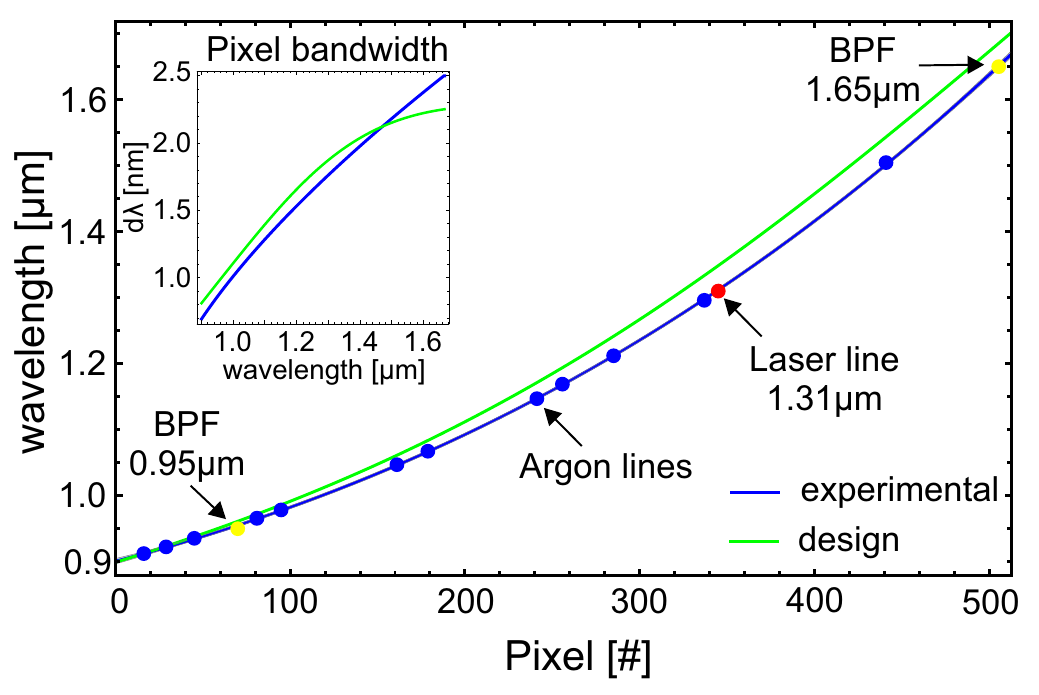}
\caption{Wavelength calibration of the InGaAs detector: Several Ar lines are identified (marked in \textit{blue}) on the InGaAs array detector. The marks for the BPF's and the alignment laser fall well in line. The \textit{blue solid line} represents a $3^{\text{rd}}$ order polynomial fit. The $3\sigma$ standard deviation of the fit model is found to be on average $\SI{0.34}{\percent}$.}
\label{fig:iDus_WLcal}
\end{figure}

The alignment of the NIR arm is performed by means of a single mode polarization-maintaining fiber optic coupled laser of central wavelength $\lambda=\SI{1.31}{\micro\metre}$ and $\Delta\lambda=\SI{10}{\nm}$ bandwidth (Thorlabs, S1FC1310PM). The central pixel value is used as the first calibration point and later reused for the absolute calibration (see Section~\ref{sec:abs_calibration}. Further calibration is based on a calibration Ar source (\textsc{Ocean Optics}) emitting low-pressure argon emission lines between $\SI{700}{nm}$ and $\SI{1.7}{\um}$. The Ar emission is trasnsported via fiber optic (Andor, A-ME-OPT-8004) and collimated by an OAP in order to transport it with folding mirrors into the spectrometer chamber. Additionally, the transmission signal of two bandpass filters (BPF) at $\SI{950}{nm}$ and $\SI{1.65}{\um}$ is included. Over all calibration points a $3^{\text{rd}}$-order polynomial fit is performed as shown in Fig.~\ref{fig:iDus_WLcal}.

\subsubsection{MCT detector}
The wavelength calibration of the MCT detector is performed using absorption lines of four different thin ($\sim \SI{10}{\um}$) Teflon films (TPX, HDPE, PP, Mylar) whose transmission spectra had been characterized over the spectral range of interest for the MCT detector beforehand by using a Fourier transform infrared (FTIR) spectrometer (Bruker, $\SI{4}{\cm^{-1}}$-resolution). These absorption lines are then measured with the MCT detector using the BB radiator as the reference LS. Over the calibration list including a $\text{CO}_\text{2}$ laser line and two BPFs at $\SI{1.65}{\um}$ and $\SI{2}{\um}$ a $3^{\text{rd}}$ order polynomial is fitted.

\begin{figure}[htbp]
\centering
\includegraphics[width=0.49\textwidth]{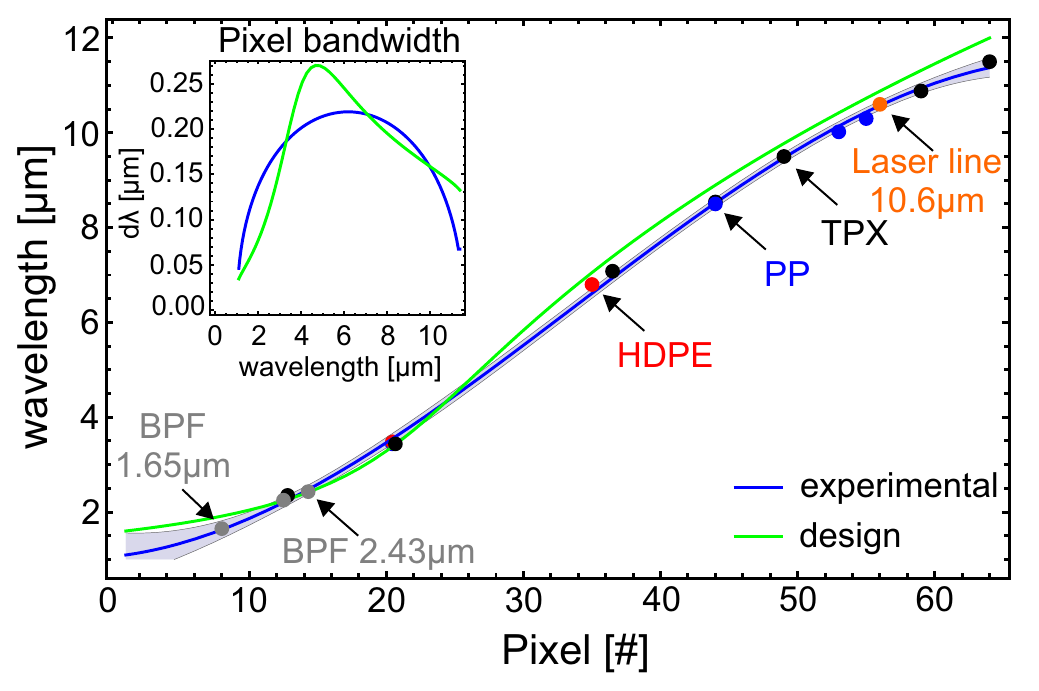}
\caption{Wavelength calibration of the MCT detector: The \textit{blue solid line} indicates a $3^{\text{rd}}$ order polynomial fit, with the light blue band representing the $3\sigma$ confidence interval of the fit.}
\label{fig:MCT_WLcal}
\end{figure}

The result of the MCT calibration, presented in Fig.~\ref{fig:MCT_WLcal}, indicates that the different absorption foils and BPFs provide a variety of calibration points which enable a relative error of only $\SI{4}{\percent}$ on average.

\subsection{Resolution and resolving power}
\begin{figure}[htbp]
\centering
\includegraphics[width=0.485\textwidth]{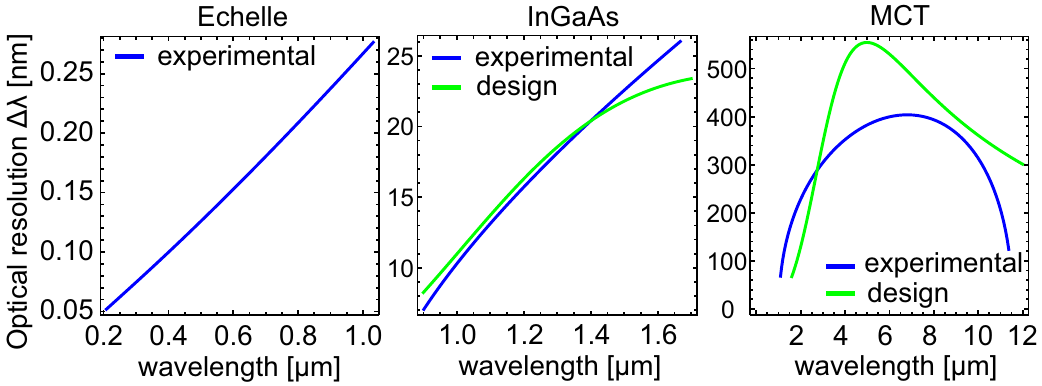}
\caption{Features the resolving power calculated for each spectrometer arm. Since the theoretical angular dispersion for the commercial echelle spectrometer is not available, only the experimental curve is presented here.}
\label{fig:resolving_power}
\end{figure}

\begin{figure}[htbp]
\centering
\includegraphics[width=0.485\textwidth]{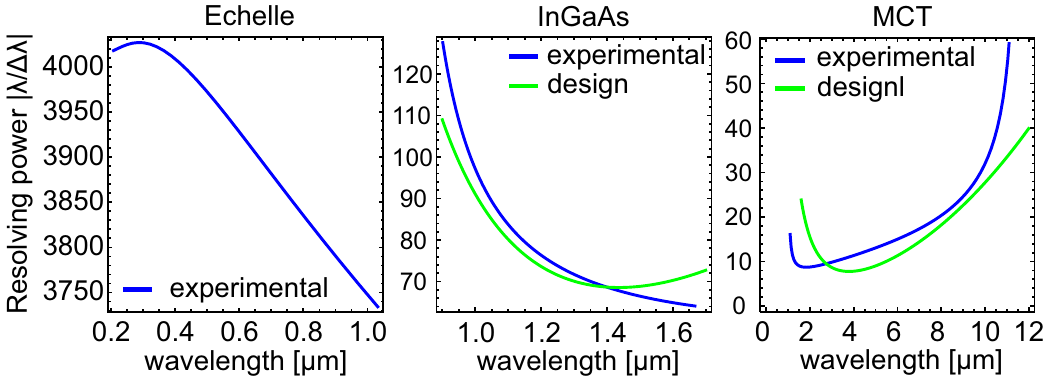}
\caption{Shows the instrument spectral resolution, individually obtained for each spectrometer arm.}
\label{fig:optical_resolution}
\end{figure}

The theoretical and experimental resolving power of each individual spectrometer arm is evaluated. Values for the theoretical angular dispersion $d\delta/d\lambda$ are obtained from Eq.~\eqref{eq:prism_design_deflection} and the experimental linear dispersion $dx / d\lambda$ follows from the wavelength fits. The latter are evaluated on a discrete grid representing the dimension of the active area of the corresponding detectors.
Figure \ref{fig:resolving_power} depicts the resolving power for each of the spectrometer arms.
Analogous to the resolving power the spectral resolution of the spectrometer is shown in Fig.~\ref{fig:optical_resolution}.
The spectral resolution increases from the MIR range to the UV range by more than 3 orders of magnitude, well matching the experimental signatures expected for frequency modulated broadband CTR spectra (see Section \ref{sec:exp_results}).

Although the smallest detectable wavelength of the echelle spectrometer amounts to $\SI{200}{\nm}$, the low throughput of the spectrometer below $\SI{250}{\nm}$ prevents an accurate relative response calibration in this range (see Section~\ref{sec:rel_calibration}). Hence, without resorting to spectral binnings to boost the signal to noise ratio, which could minimally lower this detection limit, we find the short-wavelength limit of the echelle spectrometer to be $\SI{250}{\nm}$. In contrast, the upper spectral limit of the full instrument is currently limited to $\SI{11.35}{\micro\metre}$ given by the MCT calibration fit (see Fig.~\ref{fig:MCT_WLcal}).

In summary, the free spectral range of the TR spectrometer is $\SI{250}{\nm}$ to $\SI{11.35}{\micro\metre}$ with about $\SI{100}{\nm}$ spectral overlap between the different spectrometer detectors.

\subsection{Relative response calibration} \label{sec:rel_calibration}

This first yield calibration is performed by measuring the relative spectral response of the instrument $T_{\text{LS}}(\lambda)$ in units of $[\text{counts or V}]$ to calibrated continuous light sources (LS) with $S_{\text{LS}}(\lambda)$ denoting the spectral emission characteristic of the LS in $[\text{arb. energy}]$ units. It reflects the wavelength dependent response of the detectors convoluted with the transmission efficiency of the spectrometer arms. This method enables to determine the relative efficiency of the spectrometer over a broad spectral range at once. Due to diffraction the incoherent emission of the available LSs can not be precisely collimated and transported through the spectrometer optics, which leads to a large uncertainty in the absolute yield of the LSs. Therefore, this relative calibration is normalized to unity with respect to the wavelength at which the absolute calibration will be anchored later, see Section~\ref{sec:abs_calibration}.

For the MIR and NIR ranges, a calibrated BB source (IR-508/301, \textsc{Infrared Systems}) is used. Its emission spectrum in the range from $\SI{0.5}{\um}$ up to $\SI{99}{\um}$ comes close to an ideal BB spectrum (specified $>\SI{99}{\percent}$ emissivity) described by the Planck equation. In the UV and NIR ranges a QTH lamp is used and complemented by the BB radiator and a Deuterium lamp. Here, only the calibration result from QTH is presented.

The relative calibration can thus be expressed by
\begin{align}\label{eq:relcalibration}
T_{\text{Rel}}^{\text{s,p}}(\lambda) &=\dfrac{T_{\text{LS}}^{\text{s,p}}(\lambda)}{S_{\text{LS}}(\lambda)} \cdot \dfrac{1}{R_{\text{mirrors}}^{\text{s,p}}(\lambda)} \cdot \dfrac{1}{T_{\text{polarizer}}(\lambda)}  \\
\label{eq:relcalibration_norm}
T_{\text{Rel, N}}^{\text{s,p}}(\lambda) &=T_{\text{Rel}}^{\text{s,p}}(\lambda) / T_{\text{Rel}}^{\text{s,p}}(\lambda_{\text{Cal.}}),
\end{align}
where $T_{\text{Rel}}^{\text{s,p}}$ represents the transmission of the LS through the spectrometer with respect to s- and p-polarization. $S_{\text{LS}}$ denotes the emission spectrum of the LS, which is derived from an independent calibration measurement (for QTH and Deuterium sources) or from fundamental concepts (Planck's law for the BB source).
Since alignment optics slightly alter the original LS spectrum, these contributions $R_{\text{mirrors}}^{\text{s,p}}$ as well as transmission through the polarizer $T_{\text{polarizer}}$ are explicitly taken into account.
$T_{\text{polarizer}}$ was determined with the aforementioned FTIR spectrometer. The spectral reflectivity of aluminum, gold and silver coated mirrors is taken from specifications.

Finally, the relative calibration curves are normalized to unity with respect to the value of the central pixel $ T_{\text{Rel}}^{\text{s,p}}(\lambda_{\text{Cal}})$
corresponding to the laser line $\lambda_{\text{Cal}}$ applied for the following absolute calibration.

\subsection{Absolute calibration} \label{sec:abs_calibration}

Equation \eqref{eq:relcalibration_norm} encodes information about the relative response of the spectrometer as a function of the wavelength. A subsequent energy calibration at a laser line defined reference point thus transforms the entire relative calibration into an absolute. This procedure is performed for s- and p-polarization separately.

For the EMCCD and InGaAs detectors, $\SI{532}{nm}$ and $\SI{1.31}{\um}$ CW laser sources are employed, respectively. For signal acquisition the detector exposure time is internally set to $\tau_{\text{exp}}=\SI{50}{ms}$ in order to mitigate the error in $\tau_{\text{exp}}$ originating from the intrinsic detector gating jitter and readout noise. Additionally, at the input of the echelle spectrometer a programmable mechanical shutter is implemented for minimizing further signal accumulation during readout.  Furthermore, the intensity of the laser beams is reduced by neutral density (ND) filters such that the entire dynamic range of the detector is covered. The average power $P_{\text{avg}}$ of the CW laser beam is measured by a calibrated thermal power detector (model XLP12 from Gentec) while the ND attenuators are removed from the beam in order to increase the SNR of the measurement. The involved ND filters are separately calibrated for the same laser beam and included in the final energy calculation. For a CW laser, the energy contained within a specific acquisition time $\tau_{\text{exp}}$ is thus calculated by
\begin{equation}\label{eq:laserenergy}
E_{\text{laser}} [\si{\joule}]= P_{\text{avg}} [\si{\watt}] \cdot T_{\text{ND}} \cdot \tau_{\text{exp}} [\si{\s}],
\end{equation}
where $P_{\text{avg}}$ and $T_{\text{ND}}$ denote the laser average power and the attenuation factor, respectively.

The calibration of the MCT detector requires more effort, since the detector time constant of $\SI{400}{ns}$ does not allow to perform valid measurements with pulses longer than this time constant. In the MIR range laser sources that can provide pulses below $\SI{400}{ns}$ are limited. One rather complex option is to use accelerator based FEL light sources \cite{Gabriel2000} as, e.g., provided by HZDR's light source ELBE. Here, we apply an alternative transportable method by adapting a $\text{CO}_2$ laser with an acousto-optical modulator (AOM). The AOM enables the generation of short pulses down to $\SI{100}{ns}$ with $\si{kHz}$ repetition rate.

\begin{figure}[t]
\centering
\includegraphics[width=0.47\textwidth]{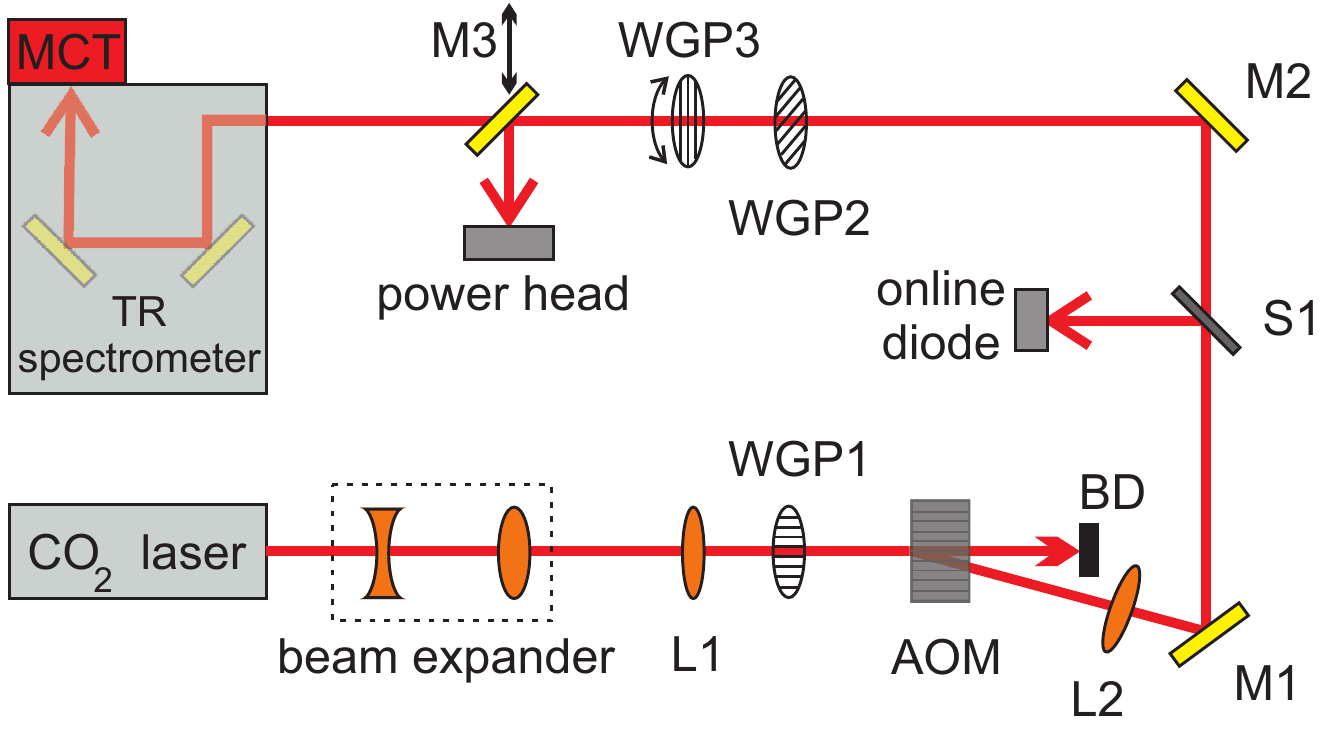}
\caption{Schemetic of the $\mathrm{CO}_2$ calibration setup: The beam expander consists of a defocusing and a focusing ZnSe lens, enlarges the beam diameter from $\sim \SI{1}{mm}$ to $\sim \SI{10}{mm}$. The beam is focused on the AOM crystal by L1 and is polarized by WGP1. The deflected pulsed beam under $\SI{161}{mrad}$ deflection angle is recollimated by L2. Following the mirror M1, a fraction of the beam is reflected by the $\SI{1}{mm}$ thick silicon wafer S1 into a fast photo-diode to monitor the laser pulse length. Transmission through the S1 passes WGP2 set at $\SI{45}{\degree}$ angle and further passes the adjustable WGP3 which is utilized to define the output polarization of the setup. The mirror M3 can be either inserted into the beam for power measurement or removed for spectrometer calibration.}
\label{fig:schemetic_AO}
\end{figure}

The $\mathrm{CO}_2$ calibration setup is depicted in Fig.~\ref{fig:schemetic_AO}.
A temperature stabilized quasi-CW $\mathrm{CO}_2$ laser with $\SI{1}{W}$ output (model L4 from manufacturer \textsc{Soliton}) is utilized.
The AOM system (model N37041-1 from ELS GmbH) uses germanium (Ge) for the IR optical interaction material with a lithium niobate transducer. The $\mathrm{CO}_2$ laser polarization axis is adjusted to the Bragg plane of the Ge-crystal (here p-polarization) using a holographic wire-grid linear polarizer (WGP1) ($\mathrm{BaF}_2$ Thorlabs WP25H-B). The AOM can be externally triggered at $f=\SI{1}{kHz}$ with an output pulse width of $\SI{200}{\ns}$ (FWHM).

The p-polarized output pulse train from the AOM setup is thereafter polarized to $\SI{45}{\degree}$ by WGP2. WGP3 then defines the final output polarization of the AOM assembly for the MCT detector calibration. The average power $P_{\text{avg}}$ of each polarization is measured using a calibrated thermal power detector (model XLP12 from Gentec).
The MCT readout electronic is synchronized to this laser pulse train.

In order to improve the SNR during the power measurement the laser pulse width is increased by a factor of 1000 to $\tau_{\mathrm{Laser}} = \SI{200}{\us}$. The pulse energy of laser pulse is calculated by $E\mathrm{[J]}  = P_{\mathrm{avg}}[W] \cdot f^{-1}[s]$ and the pulse energy during the MCT calibration measurement obeys
\begin{equation}
E_{\mathrm{laser}}\mathrm{[J]}  =\frac{\tau_{\mathrm{exp}}}{\tau_{\mathrm{Laser}}} \cdot P_{\mathrm{avg}}\mathrm{[W]} \cdot f^{-1} \mathrm{[s]}.
\label{eq:laserenergy_mct}
\end{equation}

\noindent
The absolute calibration factor is therefore given by
\begin{equation}\label{eq:absolutefactor}
F_{\mathrm{Abs}}^{s,p}=\dfrac{S_{\mathrm{laser}}}{E_{\mathrm{laser}}}.
\end{equation}
$S_{\mathrm{laser}}=\sum_{\#} S_{\mathrm{laser}}(\#)$ denotes the integrated signal from the calibration laser measured over the entire region of interest on the detector and $E_{\mathrm{laser}}$ is given by the Eq.~\eqref{eq:laserenergy} or \eqref{eq:laserenergy_mct}. The unit of $F_{\mathrm{Abs}}^{s,p}$ is depending on the detector $[\text{counts}\si{\per\joule}]$ or $[\si{\volt\per\joule}]$.

\begin{figure*}[htbp]
\centering
\includegraphics[width=0.95\textwidth]{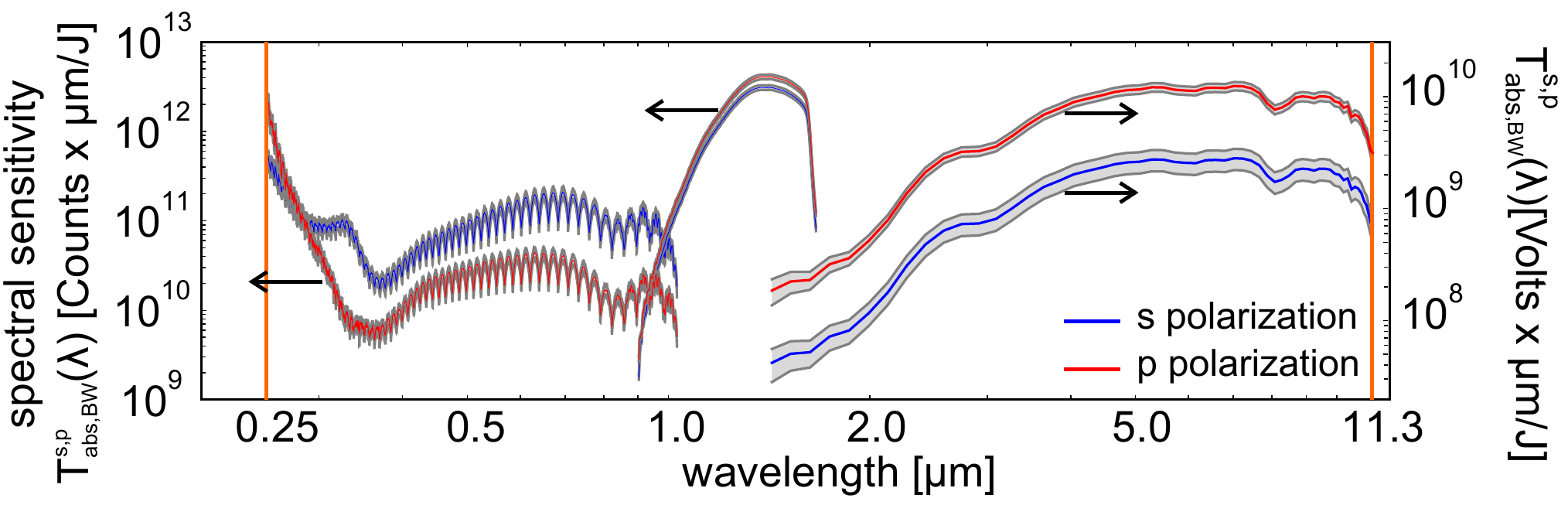}
\caption{Overview over the bandwidth-corrected absolute instrument calibration: The three sets of calibration curves represent the spectral range of the three independent detectors. Each set is presented with respect to s- and p-polarization as indicated in \textit{blue} and \textit{red}, respectively. The \textit{gray} area indicates the error contributions from the wavelength, relative and absolute calibrations. Arrows indicate the associated axes.}
\label{fig:overview_calibration_freepath}
\end{figure*}

Finally, a normalization of the absolute calibration curves by the spectral bandwidth $BW(\lambda)=\Delta\lambda$ covered by the individual detector pixels eliminates the non-linearity by non-equidistant wavelength sampling of the detectors and thus provides a consistent calibration over the entire spectrum. The bandwidth normalized absolute calibration is therefore given by
\begin{align}\label{eq:abscalibration}
T_{\mathrm{Abs}}^{s,p}(\lambda) &= T_{\mathrm{Rel, N}}^{\mathrm{s,p}}(\lambda) \cdot F_{\mathrm{Abs}}^{\mathrm{s,p}} \\
T_{\mathrm{Abs,BW}}^{s,p}(\lambda) &= T_{\mathrm{Rel, N}}^{\mathrm{s,p}}(\lambda) \cdot F_{\mathrm{Abs}}^{\mathrm{s,p}} \cdot BW(\lambda), \label{eq:abscalibration_BW}
\end{align}
where the quantity $BW(\lambda)$ indicates the discrete spectral bandwidth of the detector array for each pixel and is derived from the wavelength calibration curve of the corresponding detector (see inset on Fig.~\ref{fig:iDus_WLcal}). The absolute calibration curve $F_{\mathrm{Abs}}^{s,p}$ from Eq.~\eqref{eq:abscalibration_BW} is given in units of $\text{counts or voltage}\times\si{\um\per\joule}$ and is summarized in Fig.~\ref{fig:overview_calibration_freepath}.

The error band (depicted in gray) reflects one standard deviation and contains all error sources as follows:

\textbf{1.} The relative error from signal noise of each detector $\sigma_{S_n}$ is considered as a function of wavelength and calculated by
\begin{equation}
\sigma_{S_n}(\lambda) =\sigma_{\mathrm{Bgd}}(\lambda)/S_n(\lambda).
\end{equation}
It depends on the signal level and the noise from the environment such as temperature, readout noise but also electromagnetic pulse (EMP) events during the acquisition.

\textbf{2.} The relative error regarding the spectral bandwidth $BW(\lambda)$ of each detector element is obtained from the standard error estimated from the corresponding wavelength fit function and is calculated by
\begin{equation}
\sigma_{\Delta \lambda} = \Delta\lambda / \bar{\lambda}
\end{equation}

\textbf{3.} The error from the relative response calibration $\sigma_{\mathrm{Rel}}$ is estimated according to the absolute calibration of the LS. This error is negligible ($\approx \SI{1}{\percent}$) for the BB source and lies between $\SI{8.2}{\percent}$ to $\SI{9.4}{\percent}$ for the QTH lamp in the spectral range of echelle spectrometer.

\textbf{4.} The largest error contributions originate from the uncertainties in the absolute calibration. The error directly related to the power head calibration is $\SI{3}{\percent}$, valid for all utilized calibration lasers. In addition, the individual errors in the power measurements are: for the echelle spectrometer $\SI{7}{\percent}$, for the InGaAs $\SI{3}{\percent}$ and for the MCT heads on average $\SI{12}{\percent}$.

As will be presented in Section~\ref{sec:exp_results}, appropriate ND filters are added to the NIR and MIR arms of the TR spectrometer for exploiting the full dynamic range of the detectors. Additionally in the MIR arm, KRS-5 holographic WGPs (Thorlabs WP50H-K) are utilized pairwise, as a cross-polarizer, in order to precisely adjust the throughput to conform to the dynamic range of the MCT detector. A dedicated calibration run for TR spectrometer is performed, in which the aforementioned ND filters are employed. Hence the absolute calibration given in Eq.~\eqref{eq:abscalibration_BW} can be extended to:
\begin{eqnarray}
T_{\mathrm{Abs,BW}}^{s,p}(\lambda) =&& T_{\mathrm{Rel,N}}^{s,p}(\lambda)  \cdot  T_{\mathrm{polarizer}}^{s,p}(\lambda) \cdot T_{\mathrm{ND filter}}^{s,p}(\lambda) \times \nonumber\\
&&T_{\mathrm{beam line}}^{s,p}(\lambda) \cdot F_{\mathrm{Abs}}^{s,p} \cdot BW(\lambda), \label{eq:abscalibration_BW_actual}
\end{eqnarray}
where $T_{\mathrm{polarizer}}$, $T_{\mathrm{ND filter}}$, $T_{\mathrm{beam line}}$ denote the transmission through the cross-polarizers, transmission through the ND filters and CTR beam line efficiency, respectively. Note, that due to the radial polarization of the TR emission the beam entering into the spectrometer can be assumed to be ``unpolarized''. In practice, the polarization dependent calibration curves (Eq.~\eqref{eq:abscalibration_BW}) are averaged over s- and p-polarization in order to obtain the unpolarized calibration values (Eq.~\eqref{eq:abscalibration_BW_actual}). This approximation is valid only if the characteristic diameter of the TR single-electron point-spread-function at a given wavelength $\simeq\sqrt{2}\lambda$ \cite{Castellano1998,Artru1998,Downer2018} is significantly smaller than all delimiting (slit) apertures in the spectrometer, p.ex. smaller than the \SI{50}{\um} aperture for the echelle spectrometer.

\subsection{Detection threshold}

\begin{figure}[htbp]
\centering
\includegraphics[width=0.5\textwidth]{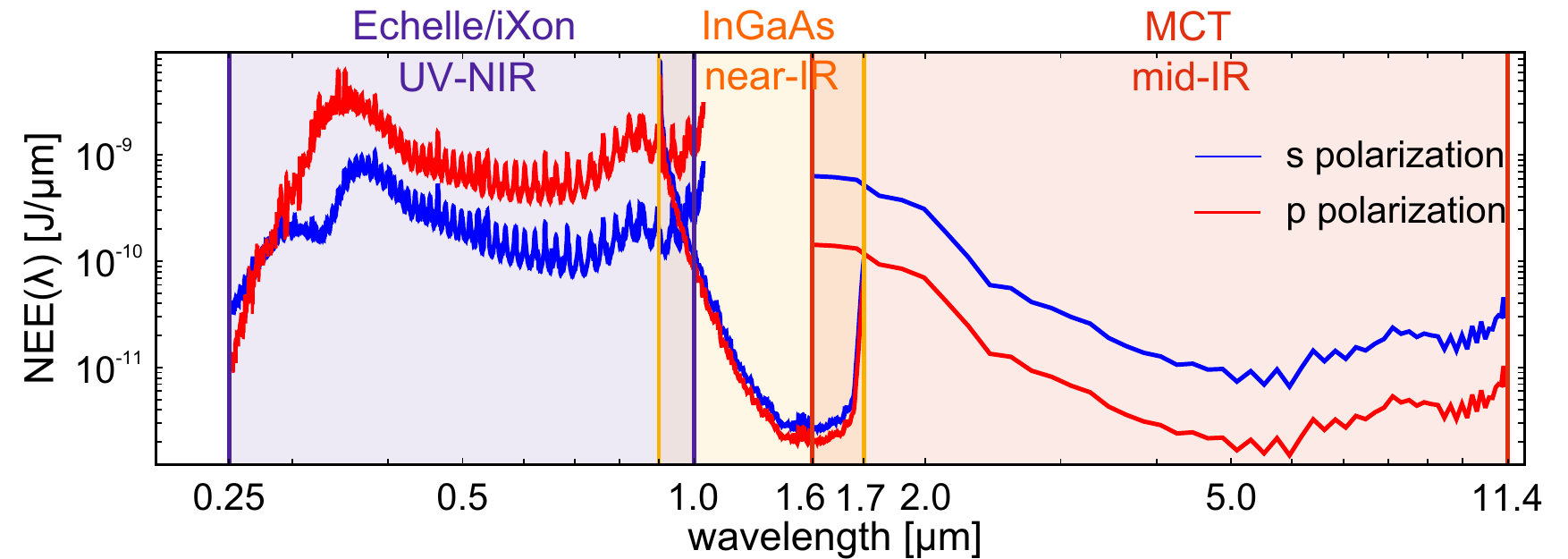}
\caption{Noise-equivalent energy (NEE) of the TR spectrometer: The NEE is estimated for the three spectral ranges and appropriate detection systems with respect to s- and p-polarization at maximum achievable resolution.}
\label{fig:overview_NEE}
\end{figure}

For the presented instrument operation over a high dynamic range is inherently coupled to high sensitivity. The latter is characterized via the noise-equivalent power (NEP), defined as the power $\phi$ radiated onto the detector which yields a signal to noise ratio (SNR) of unity~\cite{Daniels2010}.
In order to provide an estimate of the minimum energy that can be detected, the noise-equivalent energy (NEE) is introduced representing the rms noise level and is given in units of $\si{\joule\per\um}$ as
\begin{equation}
\mathrm{NEE}=\dfrac{\sigma_{\phi}}{T_{\mathrm{Abs,BW}}^{s,p}(\lambda)}.
\end{equation}
Here $\sigma_{\phi}$ denotes the standard deviation of the statistical shot-to-shot variations in background acquisitions and $T_{\text{Abs,BW}}^{s,p}$ the absolute $BW$-normalized response of the spectrometer according to Eq.~\eqref{eq:abscalibration_BW}. In the Fig.~\ref{fig:overview_NEE} the NEE is depicted for the entire spectral range of the TR spectrometer and with respect to the polarization of the incident beam.

The lowest detection threshold is achieved for the MCT and InGaAs detectors at $\SI{2}{pJ\per \um}$. The Echelle spectrometer shows a detection threshold of about $\SI{100}{pJ \per \um}$. The latter could be significantly improved by increasing its EM gain, which here is set at 30x amplification. However, an increase in the EM gain reduces the dynamic range of the detector.

The sensitivity of each spectrometer arm can be individually adapted to the dynamic range matching an application by means of attenuators or polarizers in either spectrometer arm. The NIR spectrum reveals less polarization dependency than the echelle and MIR spectra. This is due the nearly polarization independent transmission and reflection properties of optical elements of the spectrometer. In particular at $\lambda=\SI{1.31}{\um}$ this deviation is about $\SI{25}{\percent}$ from the mean.

\section{Experimental results}\label{sec:exp_results}

In order to benchmark the spectrometer under realistic experimental conditions, it was employed in a series of laser wakefield acceleration (LWFA) experiments. Following acceleration ultra-short electron bunches in the pulse duration range of 10\,fs pass through a thin metallic foil and generate transition radiation (TR) in forward direction. The emitted TR spectrum, characteristic for the temporal profile of the electron bunch, is typically broadband and covering a large dynamic range. Its full measurement enables the deconvolution of the temporal structure of the electron bunch in an independent post-processing step~\cite{Zarini2019}.

The experiments were performed at HZDR using the $\SI{100}{TW}$ DRACO laser system~\cite{Schramm2017}. It delivers $\SI{30}{fs}$ laser pulses with energies up to $\SI{3.5}{J}$ on target. A $\SI{3}{mm}$ de Laval nozzle \cite{Couperus2016} is used providing a supersonic jet composed of a $\SI{98.5}{\percent}$-He+$\SI{1.5}{\percent}$-$\mathrm{N}_2$ gas mixture. When fully ionized, an electron density of $n_{\mathrm{e}} = \SI{3.4e18}{{\cm}^{-3}}$ is generated which corresponds to a linear plasma wavelength of  $\lambda_{\mathrm{p}}=\SI{17.1}{\micro\metre}$. For electron injection and acceleration the self-truncated ionization injection (STII) scheme in the bubble-regime is exploited \cite{Couperus2017,Irman2018}. Typically, electron bunches are accelerated up to $\SI{400}{\MeV}$ with a peak charge of several hundred pC (FWHM) at relative energy spread of about $\SI{10}{\percent}$.

For the bunch duration measurement, a $\SI{5}{\micro\meter}$ steel foil located $\SI{26}{mm}$ downstream of the nozzle exit is used as the TR source. A 3-inch aluminum-coated spherical mirror (F/31) recollimates the forward TR. It is followed by five 2-inch aluminum-coated folding mirrors to transport the TR beam into the TR-spectrometer. The first off-axis parabolic mirror (AL-OAP1) (see Fig.~\ref{fig:spectrometeroverview}) images the TR source onto the $\SI{200}{\um}$ entrance slit of the instrument. The electron beam divergence is measured to about $\SI{5}{mrad}$ (rms) leading to a transverse electron beam size on the TR screen of less than $\SI{500}{\um}$. Thus, the F/31 SM leads to an image size of $\SI{60}{\um}$ in diameter fitting into the entrance slit and preventing clipping of the TR beam for many different LWFA beams. Note that this condition only guarantees that the TR beam is fully captured by the far-IR arm. For the UV/VIS and near-IR arm, subsequent slit sizes of \SI{50}{\um} and \SI{100}{\um} are smaller respectively. Thus beam size and TR beam pointing can lead to clipping and consequent signal reduction. However, we monitor the entrance slit for each shot by inserting a beam sampler in front of the echelle spectrometer, which reflects \SI{1.0}{\percent} of the visible TR into the imaging optics.

\begin{figure}[htbp]
\centering
\begin{tikzpicture}[every node/.style={inner sep=0,outer sep=0}]
    \node[anchor= north west] (img) at (0,0) {\includegraphics[width=0.5\textwidth]{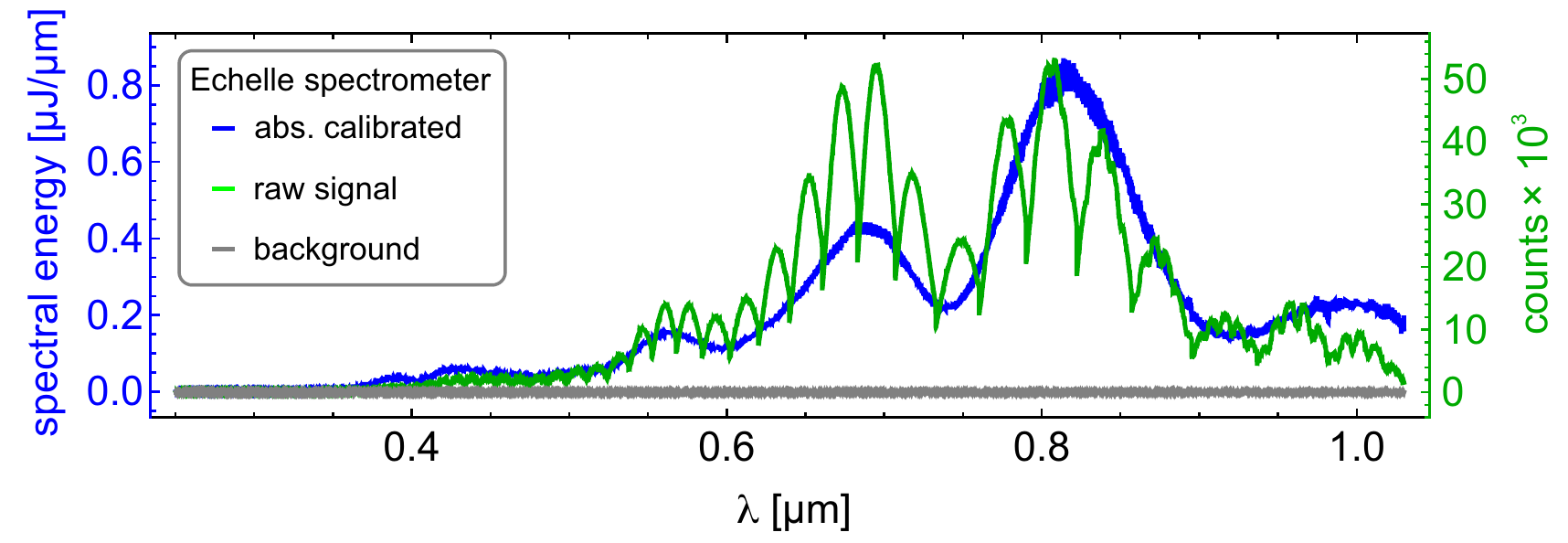}};
    \node[anchor= west] at (+3.0cm,-0.5cm) {UV/VIS};
\end{tikzpicture}
\begin{tikzpicture}[every node/.style={inner sep=0,outer sep=0}]
    \node[anchor= north west] (img) at (0,0) {\includegraphics[width=0.5\textwidth]{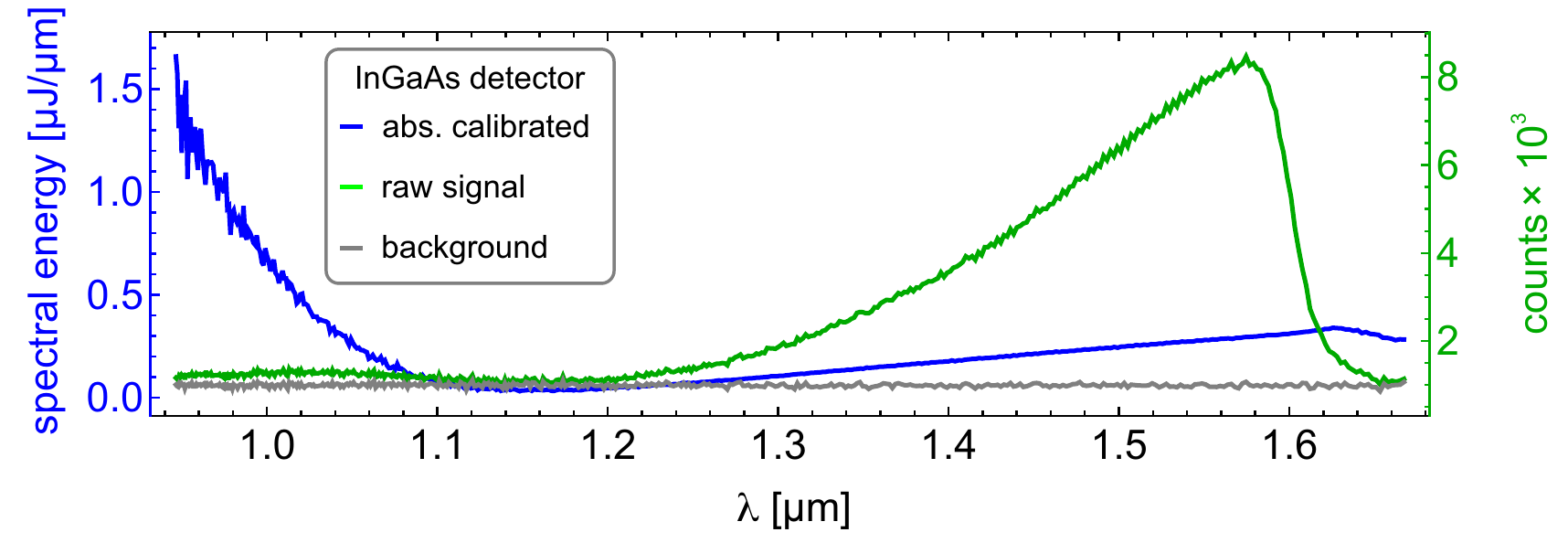}};
    \node[anchor= west] at (+4.0cm,-0.5cm) {NIR};
\end{tikzpicture}
\begin{tikzpicture}[every node/.style={inner sep=0,outer sep=0}]
    \node[anchor= north west] (img) at (0,0) {\includegraphics[width=0.5\textwidth]{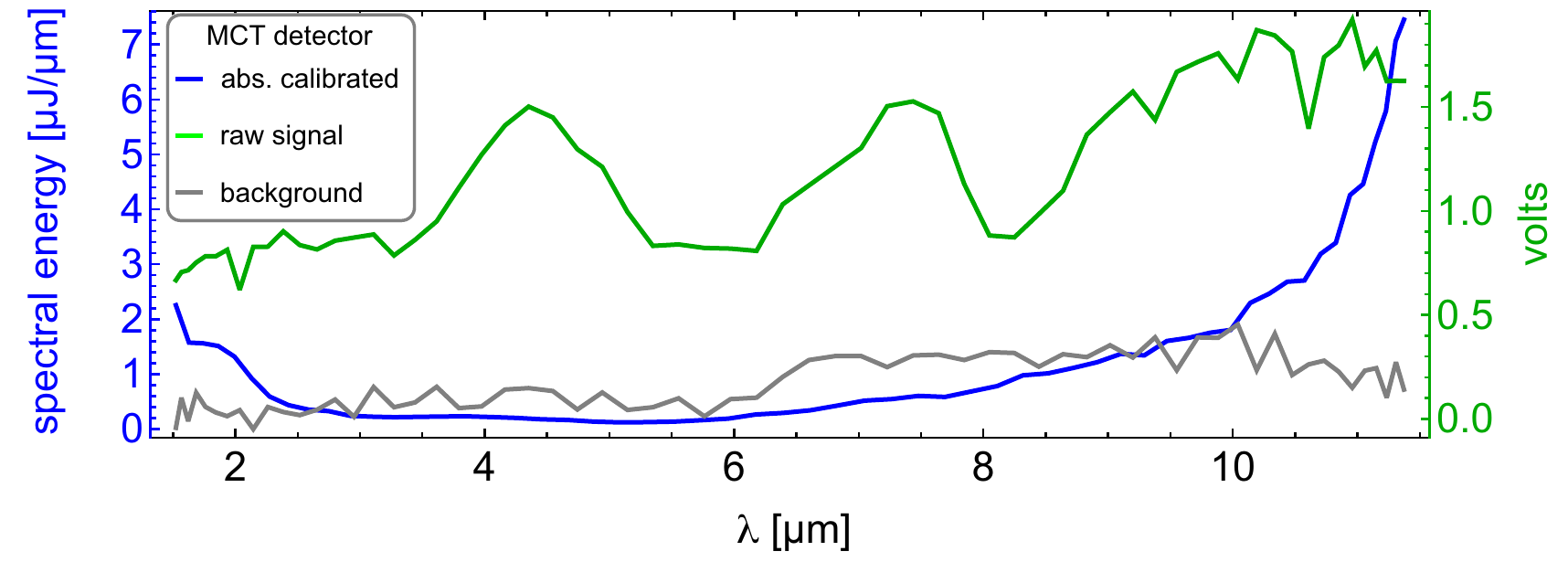}};
    \node[anchor= west] at (+4.0cm,-0.5cm) {MIR};
\end{tikzpicture}
\caption{Typical spectral measurement of coherent to incoherent TR radiation: The absolutely calibrated and bandwidth normalized partial spectra, depicted in blue, are determined using Eq.~\eqref{eq:abscalibration_BW_actual}. The green curves show the raw signal acquired by the associated detectors. The gray curves, associated also with the right axe, show the background signal acquired by removing the TR screen and SM collimator from the beam axis.}
\label{fig:res_a}
\end{figure}
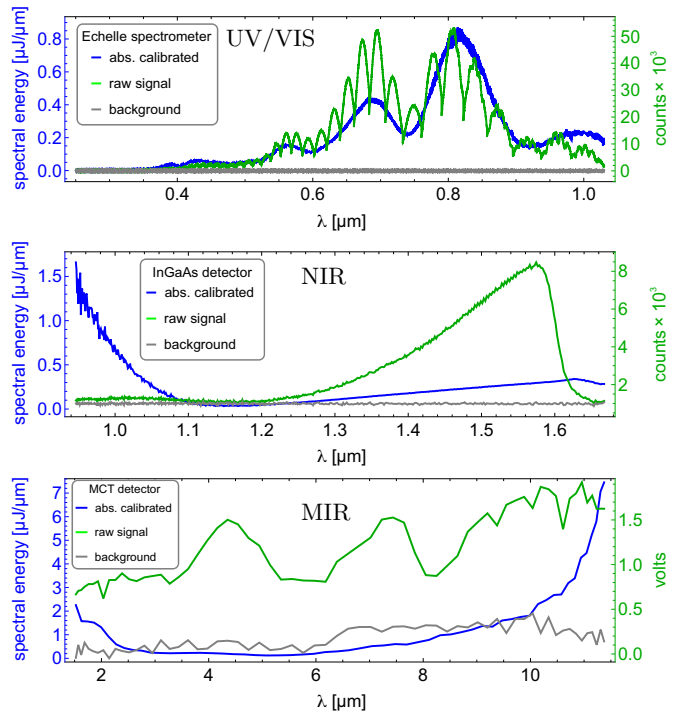

The high charge accelerated in the STII scheme leads to CTR emission of intensities up to tens of $\si{\micro\joule / \um}$ especially in the infrared range.
Fig.~\ref{fig:res_a} shows a typical set of raw and calibrated data acquired by the three spectrometer arms. Notably, the raw spectrum from the echelle spectrometer shows an extremely modulated signal as expected from varying grating efficiency. The lack of this signature in the corrected spectrum illustrates the high quality of the calibration. The corrected MCT spectrum shows a smoothly rising signal towards the long wavelength range. The latter encodes information on the overall electron bunch duration which is only accessible with a proper absolute calibration. The NIR spectrum provides the connection between the three arms. The sharp cutoff in the raw InGaAs spectrum above $\SI{1.58}{\um}$ is due to the strong decrease in its response which is partially corrected by the calibration.
Nonetheless, the overlapping regions between the arms coincide reasonably well, in particular when regarding the spectral evolution. The ratio between the measured intensities by the echelle spectrometer and InGaAs detector at $\lambda=\SI{1.0}{\um}$ is $W_{\mathrm{InGaAs}} / W_{\mathrm{Echelle}} = 4.6$.  On the other hand, the ratio between the MCT and InGaAs at $\lambda=\SI{1.63}{\um}$ is $W_{\mathrm{MCT}} / W_{\mathrm{InGaAs}} = 3.8$.

\begin{figure}[htbp]
\centering
\includegraphics[width=0.5\textwidth]{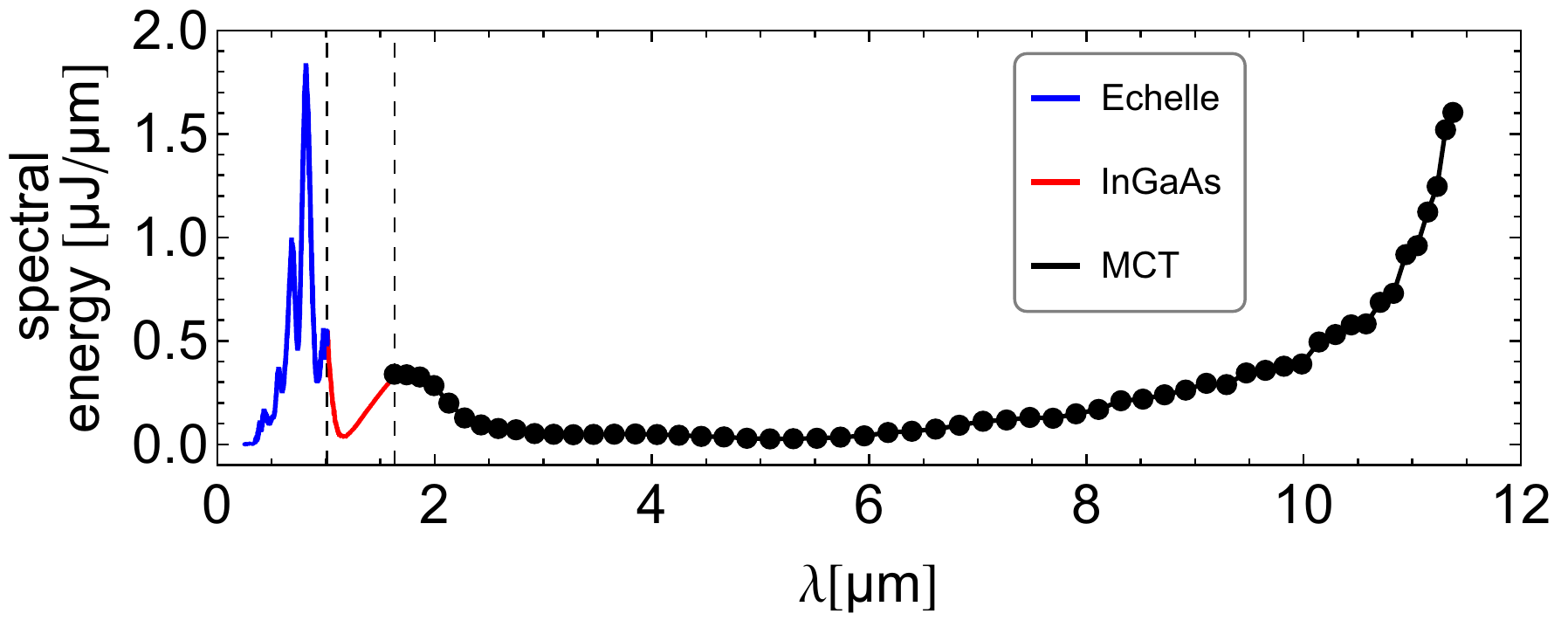}
\includegraphics[width=0.5\textwidth]{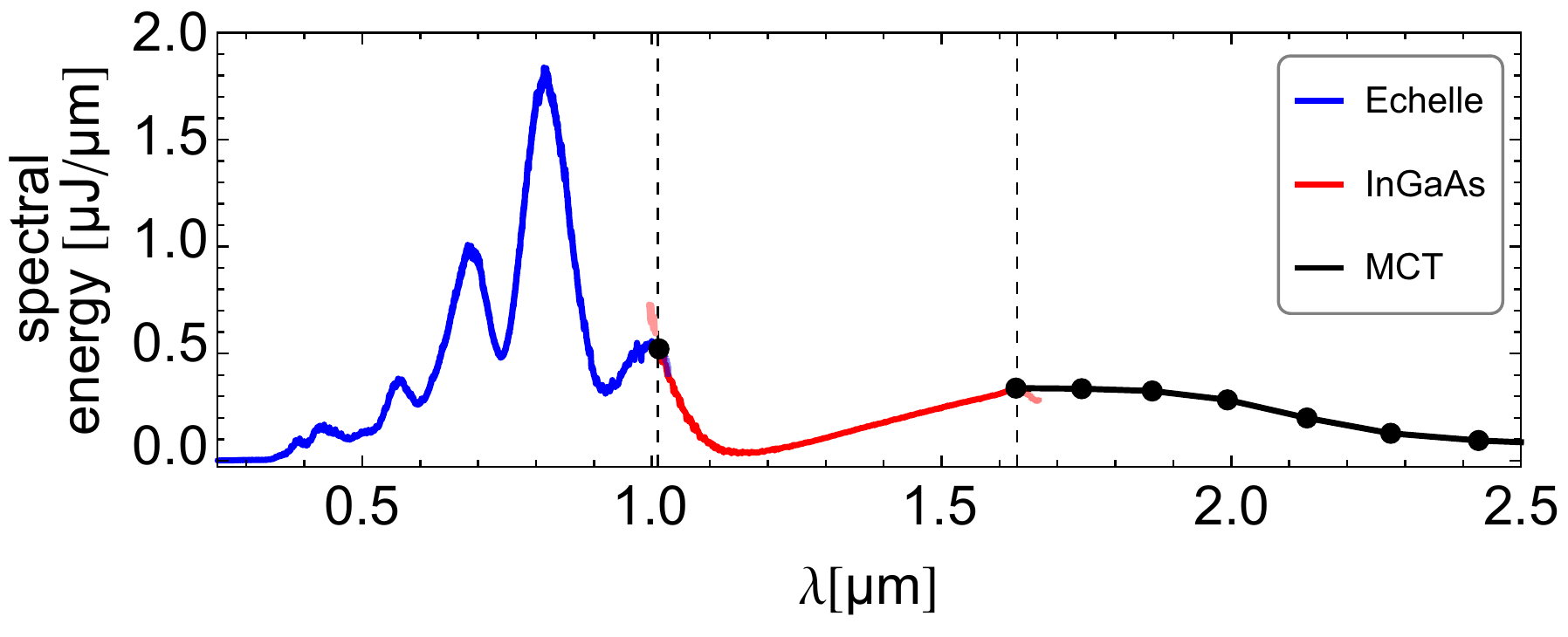}
\caption{(a) Combined measured TR spectrum: The Echelle and MCT spectra are re-scaled in their intensities according to overlap wavelengths to the InGaAs spectrum. The overlap wavelengths are $\lambda = \SI{1.0}{\um}$ and $\SI{1.63}{\um}$ for Echelle and MCT spectra respectively. (b) depicts a zoomed-in spectrum with InGaAs (pink) and Echelle data (transparent blue) beyond their designated spectral domains in regions of overlapping wavelengths.}
\label{fig:res_b}
\end{figure}

For merging the partial spectra, when full beam capture can be assumed for all detectors, the NIR spectrum serves as the reference with respect to its absolute intensities. Benefitting from the imaging geometry of the beam transport into the detector, the NIR is less sensitive to beam pointing compared to the echelle spectrum. Additionally, its calibration is more precise than the MIR spectrum due to its nearly polarization independent spectral range. Thus, the NIR and echelle spectra are re-scaled with respect to their response at the range boundaries at $\lambda_{\mathrm{MCT, min}} = \SI{1.6}{\um}$ and $\lambda_{\mathrm{Echelle, max}} = \SI{1.0}{\um}$. In the contrasting case of partial (CTR) beam capture in the NIR and UV-VIS range, but still full beam capture in the MIR, it is the MIR spectrum, which is then used as an auxiliary photometic reference. Fig.~\ref{fig:res_b} summarizes the result of such a re-scaled spectrum with the connecting wavelength values being marked by vertical dashed lines.
When comparing the responses in the overlap regions, the slopes of the UV-VIS and MIR spectral curves agree well with the NIR spectral curve progression $\leq\SI{1.0}{\um}$ and at \SI{1.6}{\um} respectively. It is worth noting that the pronounced InGaAs deviations below \SI{1.0}{\um} compared to the echelle spectrum, cannot be explained by calibration uncertainties alone. Particularly the low spectral sensitivity due to the GaAs beam splitter cut-off when combined with a low signal to noise ratio makes this spectral range of the InGaAs arm susceptible to minimal drifts in background or fine-alignment. Thus we exclude the InGaAs detector data below \SI{1.0}{\um}.

\begin{figure}[htbp]
\centering
\begin{tikzpicture}[every node/.style={inner sep=0,outer sep=0}]
    \node[anchor= north west] (img) at (0,0) {\includegraphics[width=0.5\textwidth]{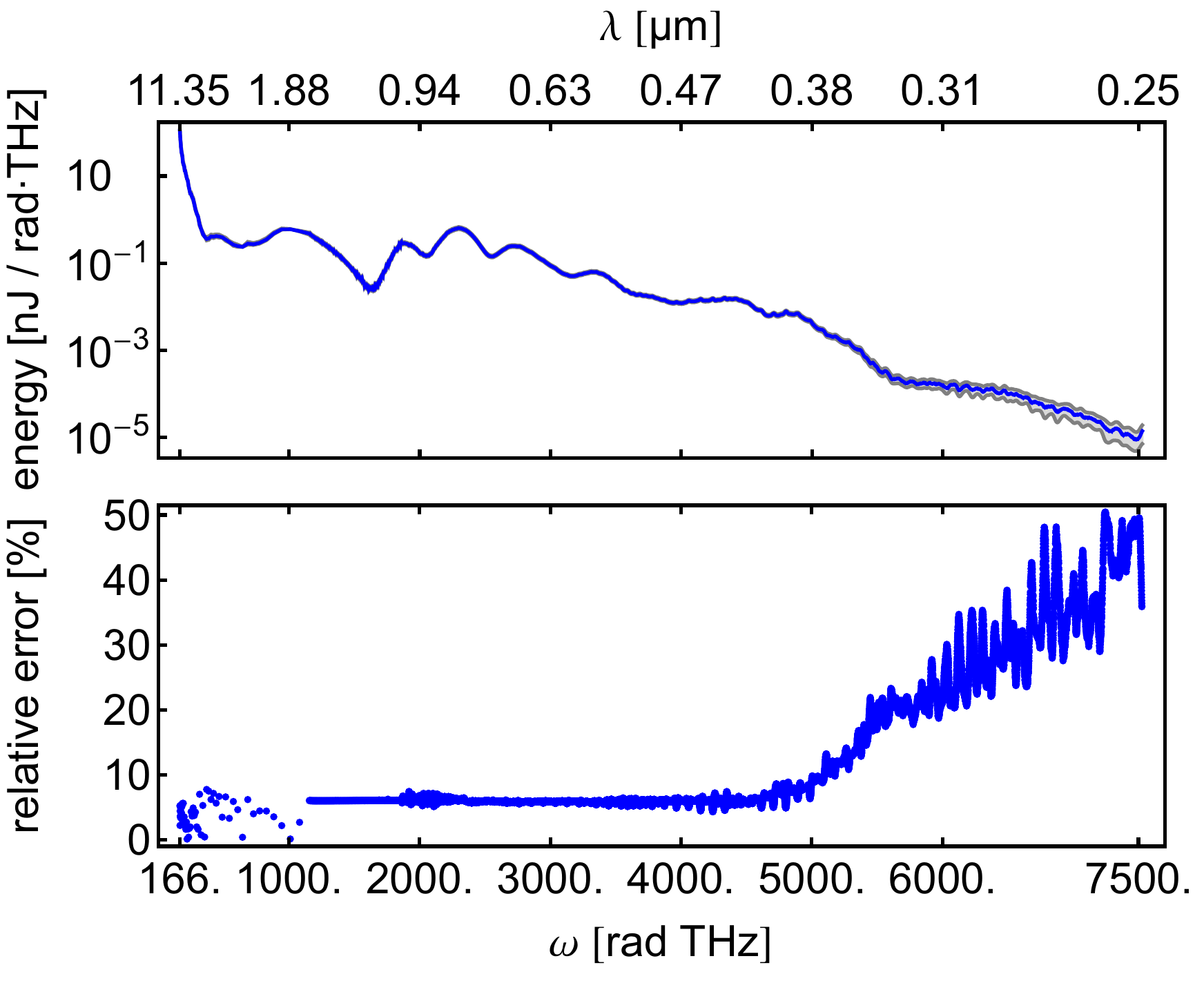}};
    \node[anchor= west] at (+1.5cm,-1.2cm) {(a)};
    \node[anchor= west] at (+1.5cm,-4.1cm) {(b)};
\end{tikzpicture}
\caption{(a) shows a typical TR spectrum in the frequency domain, combined from all three spectrometer arms. (b) depicts the corresponding relative error, dominating the UV-part of the instrument.}
\label{fig:res_c}
\end{figure}

Fig.~\ref{fig:res_c} presents the extended TR spectrum. The three spectral ranges (Fig.~\ref{fig:res_b}) are combined and subsequently returned into the frequency domain. The error band (depicted in gray) indicates a standard deviation of the mean. It is separately shown on the bottom of Fig.~\ref{fig:res_c} as a function of frequency. In the error analysis the contribution of the uncertainties of the wavelength calibration and relative calibration from each partial spectra as well as the contribution of the absolute calibration uncertainties from the NIR spectrum are taken into account.\\

\section{Conclusions}\label{sec:conclusion}

Summarizing, we presented the design, setup, full characterization, and first test of a spectrometer, which is suitable for measuring multi-octave transition radiation spectra of ultrashort electron beams in a single shot. We provide the detailed wavelength, relative response, and absolute photo-metric calibration procedure for each spectrometer arm and with respect to s- and p-polarization. This is done by using an compact ensemble of calibration light sources, situated on an optical table below the instrument.

Each of the spectrometer arms can be independently operated allowing for more flexibility in measuring broadband spectra with large variations in intensity by tailoring filter or detector configurations to typical signals. The modular setup facilitates a straightforward extension of its spectral range beyond \SI{20}{\um} (far-IR), e.g. by employing a further similar prism spectrometer to the existing MIR arm. This would be advantageous especially for measuring electron bunch durations significantly beyond \SI{20}{fs} electron durations.

The regions of spectral overlap between the 3 arms enable cross-checking the stability of calibration and alerting to potential changes in alignment. Due to its large angle of acceptance the mid-IR arm can act as an auxiliary photometric reference for the other arms in case their signals are partially clipped in secondary apertures. For monitoring beam transmission through the spectrometer entrance slit, additionally we installed a camera picking up and imaging \SI{1}{\percent} of the signal before the echelle spectrometer, hence enabling diagnostics on eventual beam pointing issues.

The resulting spectrometer enables quantitative studies of highly-modulated spectra over a broad spectral range (5.5 octaves, $\SI{250}{\nano\metre}$ to $\SI{11.35}{\micro \metre}$) at single shot. This is demonstrated by measuring a broadband TR spectrum with a dynamic range of more than 7~orders of magnitude. This provides the basis for measuring coherent transition radiation spectra from ultrashort laser-wakefield accelerated electron bunches for longitudinal density profile reconstruction \cite{Zarini2019} simultaneously resolving features from the sub-fs to few-\SI{10}{fs} scale. The systematic results on the electron bunch duration measurement are currently evaluated and will be published separately.

\section{Acknowledgments}
This work was partially funded by EUCARD2 under Grand Agreement number 312453. This project is fully supported by the Helmholtz association under program Matter and Technology, topic Accelerator Research and Development. We wish to thank M. Sobiella, A. Winter, M. Klopf, S. Grams, C. Eisenmann and M. Werner  for their technical support.

\end{document}